\documentclass[aps,prd,reprint,twocolumn,superscriptaddress,longbibliography,nofootinbib,floatfix,showpacs]{revtex4-2}
\usepackage{epsfig}
\usepackage{amsmath,amssymb,amsfonts}
\usepackage{hyperref}
\usepackage{mathrsfs}
\usepackage{bbm}
\usepackage{slashed}
\usepackage{graphicx}
\usepackage{verbatim}
\usepackage{bm}
\usepackage[usenames]{color}

\definecolor{darkgreen}{rgb}{0.2,0.6,0}

\newcommand{\be}{\begin{equation}}
\newcommand{\ee}{\end{equation}}
\newcommand{\bw}{\begin{widetext}}
\newcommand{\ew}{\end{widetext}}
\newcommand{\bi}{\begin{itemize}}
\newcommand{\ei}{\end{itemize}}
\newcommand{\ud}{\mathrm{d}}

\newcommand{\LCm}{{\scriptscriptstyle -}} 
\newcommand{\LCp}{{\scriptscriptstyle +}}
\newcommand{\LCpm}{{\scriptscriptstyle \pm}}

\newcommand{\LCperp}{{\scriptscriptstyle \perp}}

\usepackage[T1]{fontenc} \usepackage[latin1]{inputenc}

\begin{document}

\title{Worldline instantons for nonperturbative particle production by space and time dependent gravitational fields}

\author{Philip Semr\'en}
\email{philip.semren@umu.se}

\author{Greger Torgrimsson}
\email{greger.torgrimsson@umu.se}
\affiliation{Department of Physics, Ume{\aa} University, SE-901 87 Ume{\aa}, Sweden}

\begin{abstract}

We develop a worldline-instanton approach for calculating the momentum spectrum of particles produced by gravitational fields which depend on both space and time. The instantons are open. The middle part is complex and describes the formation region, while the ends describe the trajectories of asymptotic particles.   

\end{abstract}
\maketitle

\section{Introduction}

Particles can be produced nonperturbatively by gravitational fields, $g_{\mu\nu}(x^\rho)$, such as in inflation~\cite{Parker:2025jef} or by black holes~\cite{Hawking:1975vcx}. The pair-production probability, $P(g_{\mu\nu})$, can be obtained by solving e.g. the Klein-Gordon equation and calculating the Bogoliubov coefficients~\cite{Parker:2025jef}. In principle, this gives the exact dependence of $P$ on $g_{\mu\nu}$. In practice, it can be challenging to solve the Klein-Gordon equation in background fields that depend on more than one space-time coordinate.    
In this paper, we will show how to find a semi-classical approximation of $P(g_{\mu\nu})$ using a worldline-instanton approach. In comparison to other semi-classical methods~\cite{Srinivasan:1998ty,Shankaranarayanan:2000qv,Parikh:1999mf,Vanzo:2011wq}, we will show that the instanton approach is particularly useful for obtaining the momentum spectrum, $P[g_{\mu\nu}(t,x); p_\mu,p'_\mu]$, where $p_\mu$ and $p'_\mu$ are the momenta of the produced particles, for metrics which depend on more than one space-time coordinate. It also gives an intuitive picture that is not obvious from the solution to the Klein-Gordon equation.    

Instantons were first used in~\cite{Affleck:1981bma} to calculate an all-orders-in-$\alpha$ result for Schwinger pair production by a constant electric field. In~\cite{Dunne:2005sx,Dunne:2006st} it was shown that instantons provide a powerful method for studying Schwinger pair production in inhomogeneous fields, $F_{\mu\nu}(x^\rho)$, even for fields that depend on all space-time coordinates~\cite{Dunne:2006ur,Schneider:2018huk}. Those papers considered {\it closed} worldline instantons, i.e. $x^\mu(\tau_{\rm start})=x^\mu(\tau_{\rm end})$, which give the pair-production probability integrated over all momenta and summed over spins. For fields which only depend on time, it is possible to obtain the momentum spectrum from closed loops~\cite{Dumlu:2011cc}, but that is not possible in general. 

For time and space-dependent fields, the spectrum can be obtained using {\it open} instantons~\cite{DegliEsposti:2023qqu,DegliEsposti:2022yqw,DegliEsposti:2024upq}, where the ends of the instanton describe the asymptotic trajectories of the electron and positron\footnote{Pair production by constant fields has been studied using open worldlines in~\cite{Barut:1989mc,Rajeev:2021zae}.} and how they connect with the asymptotic particle states. The open-instanton approach can also be used for other processes, such as $\gamma\to e^\LCp e^\LCm$~\cite{DegliEsposti:2023fbv}, $e^\LCm\to e^\LCm\gamma$ (with a hard photon)~\cite{DegliEsposti:2021its}, or $e^\LCm\to e^\LCm e^\LCm e^\LCp$~\cite{DegliEsposti:2024rjw}. In this paper, we will generalize this open-instanton approach from $F_{\mu\nu}(x^\rho)$ to $g_{\mu\nu}(x^\rho)$. 

For black holes that only depend on one coordinate, $r$, it has been shown previously, with other semi-classical approaches~\cite{Kim:2007ep,Srinivasan:1998ty,Shankaranarayanan:2000qv,Angheben:2005rm,Moss:2023kah}, that Hawking's result can be obtained from integrals over $r$ performed using the Cauchy residue theorem, with a pole at the horizon, $r=r_H$. For 1D fields, the worldline approach could produce the same integrals as in~\cite{Kim:2007ep,Srinivasan:1998ty,Shankaranarayanan:2000qv,Angheben:2005rm,Moss:2023kah} with a parameterization of the worldline using $r$ instead of proper time $\tau$. In the Schwinger case, the corresponding transformation from instanton to WKB integrals was shown already in~\cite{Dunne:2005sx,Dunne:2006st}, while for Hawking radiation, it has only been done recently\footnote{Worldline path integrals were considered in~\cite{Hartle:1976tp,Chitre:1977ip} though.}~\cite{Lin:2024jug}. 
However, the 1D case (e.g. $E(t)$, $E(z)$, $g_{\mu\nu}(t)$ or $g_{\mu\nu}(r)$) is special and allows for simplifications that are not possible for $F_{\mu\nu}(t,z)$ or $g_{\mu\nu}(t,r)$. Indeed, the fact that one can change parameterization from $\tau$ to the single nontrivial coordinate (e.g. $r$, $t$ or $z$) means that one actually does not need to find an explicit instanton solution, because it becomes irrelevant due to the freedom to make contour deformations in the integrals over that coordinate. However, for 2D fields (e.g. $E(t,z)$ or $g_{\mu\nu}(t,r)$), one does need to find the instantons.  
Thus, it is only by considering 2D fields (e.g. $E(t,z)$ or $g_{\mu\nu}(t,r)$) that one sees the increased complexity but also the power of the worldline approach.

\section{Derivation}

The worldline representation of the propagator in arbitrary $F_{\mu\nu}(x^\rho)$ and $g_{\mu\nu}(x^\rho)$ is given by a Feynman path integral,
\be
G(x_\LCp,x_\LCm)=\int_0^\infty \ud T\int_{q(0)=x_\LCm}^{q(1)=x_\LCp}\mathcal{D}q\dots e^{-iS} \;,
\ee
where the ellipses stand for terms which will not affect the exponential part of the probability, and the exponential part of the integrand is given by~\cite{Parker:1979mf,Bekenstein:1981xe,Bastianelli:1998jm,Bastianelli:1998jb,Bastianelli:2000nm,Schubert:2001he,Bastianelli:2002fv,Bastianelli:2004zp,Edwards:2019eby} 
\be\label{worldlineAction}
S=\int_0^1\ud\tau\left(\frac{g_{\mu\nu}(q)\dot{q}^\mu\dot{q}^\nu}{2T}+A_\mu(q)\dot{q}^\mu+\frac{T}{2}[m^2+\xi R(q)]\right) \;,
\ee
where $T$ is the total proper time, $\tau$ is a normalized propertime that parametrizes the worldline $q^\mu(\tau)$, $\dot{q}=\ud q/\ud\tau$, and the path integral is a sum over open worldlines. We use units with $m=1$\footnote{In this paper we consider fields which do not depend on $y$ and $z$, so the transverse momentum components, $p_\LCperp=(p_y,p_z)$, are conserved. The dependence on $p_\LCperp$ is simply $m\to m_\LCperp=\sqrt{1+p_\LCperp^2}$.}. The massless limit can be taken by first reintroducing $m$ in the formulas below using dimensional analysis. $R$ is the Ricci scalar.

We will perform the integrals with the saddle-point method. To justify this, we consider fields with a typical length or time scale given by $1/\omega$. For example, a pulse proportional to $\text{sech}^2(\omega t)$. We expand to leading order (LO) in $\omega\ll1$, which means that there should not be any nontrivial dependence on $\omega$. We can see this explicitly, right from the start, by rescaling $q^\mu\to q^\mu/\omega$ and $T\to T/\omega$, which gives
\be
S=\frac{1}{\omega}(\text{terms without }\omega)+\omega\frac{T}{2}\int \xi R \;,
\ee
where we have used $R\sim\partial^2g\sim\omega^2$.
Thus, all the terms in~\eqref{worldlineAction} are proportional to $1/\omega$, except for the last term, which is negligible. For spin-$1/2$ particles we also have a term in the exponent proportional to $T\sigma^{\mu\nu}F_{\mu\nu}\sim\omega^0$, which is not negligible, but should be thought of as a pre-exponential term because it is $\mathcal{O}(\omega^0)$. Thus, neither the spin nor the $\xi R$ term affect the saddle-point equations or the exponential part of the final result of the probability.  

The probability amplitude, $M$, is given by the LSZ amputation of the propagator. If both particles end up in Minkowski without any electromagnetic fields (assuming also that $A_\mu(q^\nu)\to0$), then the asymptotic states are just plane waves, $\propto e^{ipx}$, and we have
\be\label{LSZ}
M=\lim_{t_\LCpm\to\infty}\int\ud^3 x_\LCp\ud^3 x_\LCm\, e^{ip x_\LCp+ip' x_\LCm}... G(x_\LCp,x_\LCm)... \;,
\ee
where the ellipses stand for parts of the formula which will not affect the exponential part of the probability, and $p_\mu$ ($p_\mu'$) is the momentum of the particle (antiparticle).

The idea is to perform all (nontrivial) integrals with the saddle-point method. The saddle ``points'' for the path integral are called worldline instantons. The equations determining the saddle points are: the Lorentz-force/geodesic equation
\be\label{geodesicEq}
\ddot{x}^\mu+\Gamma^\mu_{\nu\sigma}\dot{x}^\nu\dot{x}^\sigma=T{F^\mu}_\nu\dot{x}^\nu \;,
\ee
an on-shell condition
\be\label{TsaddleEq}
T^2=\int\ud\tau\, g_{\mu\nu}\dot{q}^\mu\dot{q}^\nu=q_{\mu\nu}\dot{q}^\mu\dot{q}^\nu \;,
\ee
and conditions for the asymptotic (and observable) momenta
\be\label{asymptoticMomentum}
\dot{q}^\mu(1)=T p^\mu \qquad 
\dot{q}^\mu(0)=-T p^{\prime\mu} \;.
\ee

We could use this formalism to study pair production in combinations of gravitational and electromagnetic fields, e.g. de Sitter plus constant electric field~\cite{Garriga:1993fh,Kim:2008xv,Stahl:2015gaa,Bavarsad:2016cxh}, but in this paper we will focus on purely gravitational fields. Eq.~\eqref{TsaddleEq} is an implicit equation for $T$, since $T$ also appears in $q^\mu(\tau)$. However, for the exponential part of the probability, we do not actually need to find $T$. By changing variable from $\tau$ to $u=T(\tau-\tau_0)$, where $\tau_0$ is a constant, $T$ drops out of all equations. We choose $\tau_0\sim1/2$ so that $u=0$ is in the middle of the instanton. We change notation so from now on $\dot{q}=\ud q/\ud u$ and by proper time we will refer to $u$. 
The on-shell condition is then given by
\be\label{TsaddleEq2}
g_{\mu\nu}\dot{q}^\mu\dot{q}^\nu=1 \;,
\ee
and the condition for the momenta 
\be\label{asymptoticMomentum2}
\dot{q}^\mu(u_1)=p^\mu \qquad 
\dot{q}^\mu(u_0)=-p^{\prime\mu} \;,
\ee
where $\text{Re }u_1\gg1$ and $\text{Re }u_0\ll-1$. We have a great deal of freedom in choosing different complex $u$ contours, but it is in general not possible for $u$ to follow the real line along the entire worldline. It also does not make sense to try to rotate to some sort of Euclidean version of~\eqref{worldlineAction}.  
In fact, as we will show, much of the work to apply these methods involves navigating in the complex $u$ plane to make sure that one stays on the correct side of branch points and singularities, which move around as we change the field parameters and/or the momenta.   

Plugging the solution into~\eqref{worldlineAction} gives
\be\label{worldlineAction2}
S=\int_{u_0}^{u_1}\ud u\, g_{\mu\nu}(q)\dot{q}^\mu\dot{q}^\nu=u_1-u_0 \;.
\ee
Squaring the amplitude gives the final result for the exponential part of the probability, $P\sim |M|^2\sim e^{-\mathcal{A}}$, where
\be\label{generalExpFin0}
\mathcal{A}=2\text{Im}[pq(u_1)+p'q(u_0)-u_1+u_0] \;.
\ee
By differentiating~\eqref{generalExpFin0} with respect to $u_0$ or $u_1$, and using~\eqref{asymptoticMomentum2}, we see that the result does not depend on the arbitrary choice of $u_0$ and $u_1$, as long as they are sufficiently large so that the instanton has left the field, $\ddot{q}^\mu(u_{0,1})\approx0$. 

While~\eqref{generalExpFin0} is finite for large $u_{0,1}$, we will derive an alternative form of $\mathcal{A}$, which makes this explicit. We first make a partial integration,
\be
S=g_{\mu\nu}(q)\dot{q}^\mu q^\nu\big|_{u_0}^{u_1}-\int_{u_0}^{u_1}\ud u\, q^\mu\partial_u[g_{\mu\nu}(q)\dot{q}^\nu] \;.
\ee
Since we assume that the metric is asymptotically Minkowski, $g_{\mu\nu}\to\eta_{\mu\nu}$ as $u\to u_0$ or $u\to u_1$, the boundary terms exactly cancel against the asymptotic states,
\be\label{expEfterPartialInt}
ip x_\LCp+ip' x_\LCm-iS=i\int_{u_0}^{u_1}\ud u\, q^\mu\partial_u[g_{\mu\nu}(q)\dot{q}^\nu] \;.
\ee
In the resulting integral we can take $u_{0,1}\to\pm\infty$. Note that all the asymptotic parameters, $T$, $t_\LCpm$ or $u_{0,1}$, have dropped out. We can further simplify~\eqref{expEfterPartialInt} by using the geodesic equation, which gives   
\be\label{generalExpFin}
\mathcal{A}=\text{Im}\int_{u_0}^{u_1}\ud u\, q^\mu g_{\nu\sigma,\mu}\dot{q}^\nu\dot{q}^\sigma \;.
\ee
Eq.~\eqref{generalExpFin} is equivalent to~\eqref{generalExpFin0}. Evaluating both allows us to check the precision of the result.

To illustrate these methods, we will focus on 2D fields that depend on time and one spatial coordinate,
\be\label{abc}
\ud s^2=a(t,x)\ud t^2+2c(t,x)\ud t\ud x-b(t,x)\ud x^2 \;.
\ee

The asymptotic conditions in~\eqref{asymptoticMomentum2} allow us to consider arbitrary values of the final momenta. The spectrum is peaked at the saddle-point values, $p_s$ and $p_s'$, which are determined by
\be
\frac{\partial\mathcal{A}}{\partial p_x}=
\frac{\partial\mathcal{A}}{\partial p'_x}=0 \;.
\ee
Eq.~\eqref{generalExpFin} has a nontrivial dependence on $p$ and $p'$ via the instanton $q^\mu(p,p')$, but the first-order momentum derivatives are actually simple to compute. We just have to express $\mathcal{A}$ in its original form~\eqref{LSZ}, where all integration variables are replaced by their saddle-point values but were we have not performed the simplifications that lead to~\eqref{generalExpFin}, and then use
\be\label{udToPartial}
\frac{\ud}{\ud p_x}\mathcal{A}[p_x,Q_s(p_x)]=\frac{\partial}{\partial p_x}\mathcal{A}[p_x,Q_s(p_x)] \;,
\ee
where $Q$ stands for all the integration variables and $Q_s(p_x)$ their saddle-point values. Thus,  
\be\label{dAdp}
\frac{\partial\mathcal{A}}{\partial p_x}=2\text{Im}\left[x(u_1)+\frac{p_x}{p_0}t(u_1)\right]
=2\text{Im}\left[x-\frac{\dot{x}}{\dot{t}}t\right](u_1)
\ee
and
\be\label{dAdpp}
\frac{\partial\mathcal{A}}{\partial p'_x}=2\text{Im}\left[x(u_0)+\frac{p'_x}{p'_0}t(u_0)\right]
=2\text{Im}\left[x-\frac{\dot{x}}{\dot{t}}t\right](u_0) \;.
\ee
The instanton for $p=p_s$ and $p'=p_s'$ is therefore determined by
\be\label{saddleCondition}
\text{Im }x=\frac{\dot{x}}{\dot{t}}\text{Im} t
\qquad
\text{Im }\dot{x}=0
\qquad
\text{for } u=u_0 \text{ or } u_1 \;.
\ee
The conditions in~\eqref{saddleCondition} allow us to find the saddle-point instanton without using~\eqref{asymptoticMomentum2}. This is the same idea as in the QED case~\cite{DegliEsposti:2022yqw}. $u_0$ and $u_1$ should be large enough so that~\eqref{saddleCondition} should remain true if we continue past $u_0$ and $u_1$.      
For $p\ne p_s$ and $p'\ne p_s'$, the instanton does not satisfy~\eqref{saddleCondition}, and then $\text{Im }x(u_{0,1})$ and $\text{Im }t(u_{0,1})$ cannot both be zero. If we choose a $u$ contour such that $\text{Im }u=\text{Im }u_0=\text{const.}$ and $\text{Im }u=\text{Im }u_1=\text{const.}$ as $\text{Re }u\to-\infty$ or $\text{Re }u\to\infty$, then, since $\text{Im }\dot{q}^\mu(u_{0,1})=0$, $|\text{Re }q^\mu(u)|\to\infty$ while $\text{Im }q^\mu(u)\to\text{const.}$, so $\text{Im }q$ will become small compared to $\text{Re }q$. It is not a problem that the instanton does not become real asymptotically, because we are only asking for the probability to observe the particles with certain asymptotic momenta, not that the particles are observed in some given region of space (although that could be an interesting problem).          

As we will demonstrate below, it can be quite nontrivial to find instantons. It is a good idea to start with a simple metric, for which it is relatively easy to find instantons, and then do a numerical continuation towards more nontrivial metrics. We will demonstrate this idea using two types of starting points: $\{a,b,c\}=\{1,b(t),0\}$ in Sec~\ref{inflation section} and $\{a,b,c\}=\{1-v^2(x),1,v(x)\}$ in Sec~\ref{acoustic section}.

\begin{figure}
    \centering
    \includegraphics[width=\linewidth]{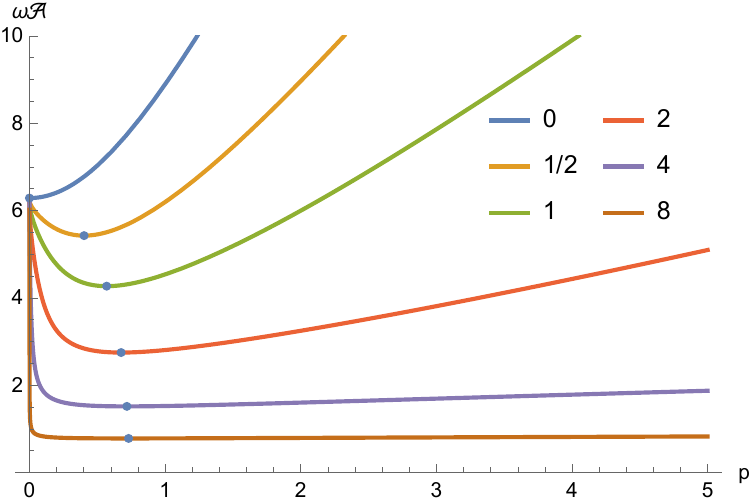}
    \caption{Results for~\eqref{ASauterSauter} with $\beta=0$, and $\alpha=0,1/2,1,2,4,8$. The dots show the location of the saddle points. The asymptotic momentum is $\dot{x}(\infty)=pe^{-2\alpha}$.}
    \label{fig:ASauterTime}
\end{figure}

\begin{figure*}
    \centering
    \includegraphics[width=0.24\linewidth]{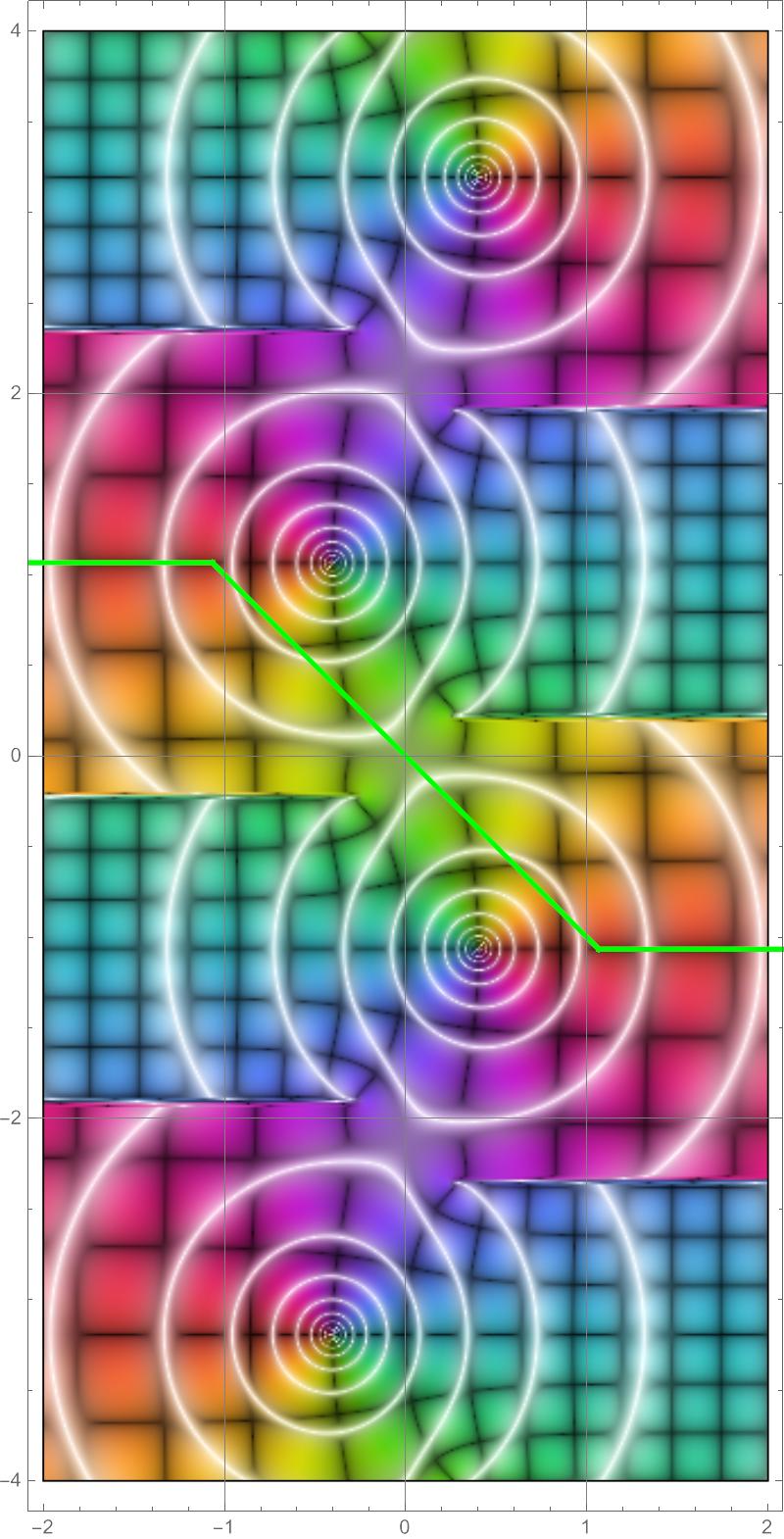}
    \includegraphics[width=0.24\linewidth]{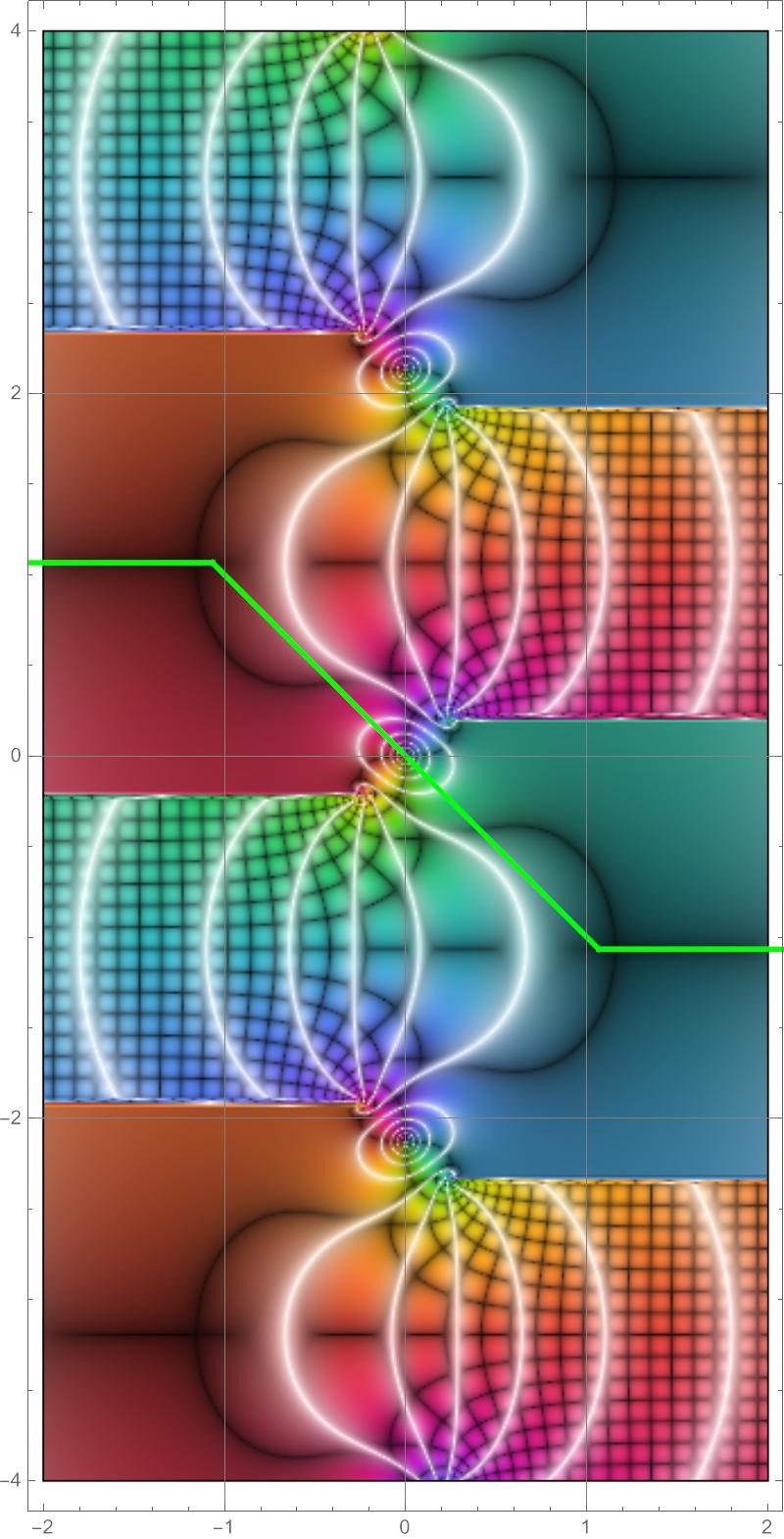}
    \includegraphics[width=0.24\linewidth]{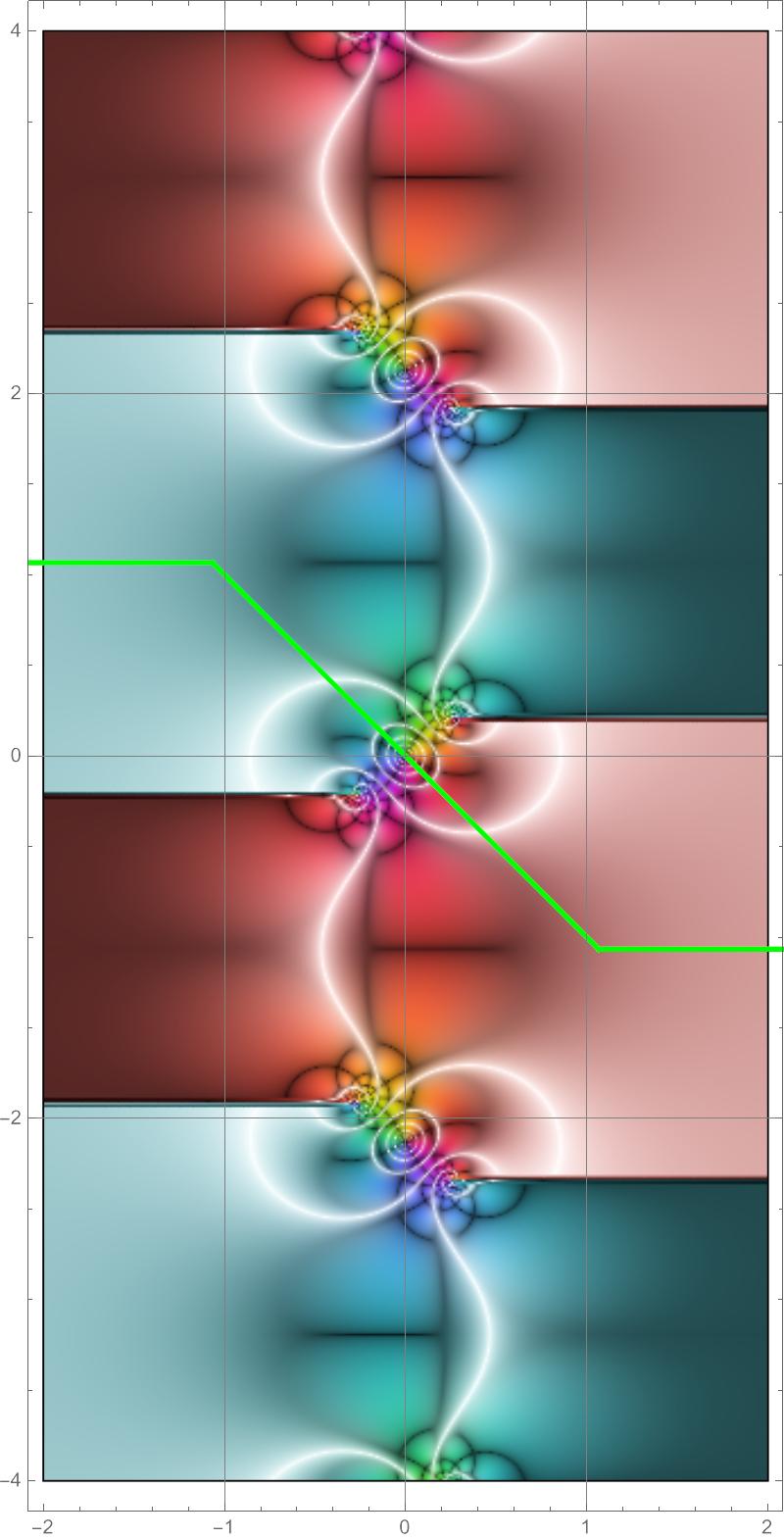}
    \includegraphics[width=0.24\linewidth]{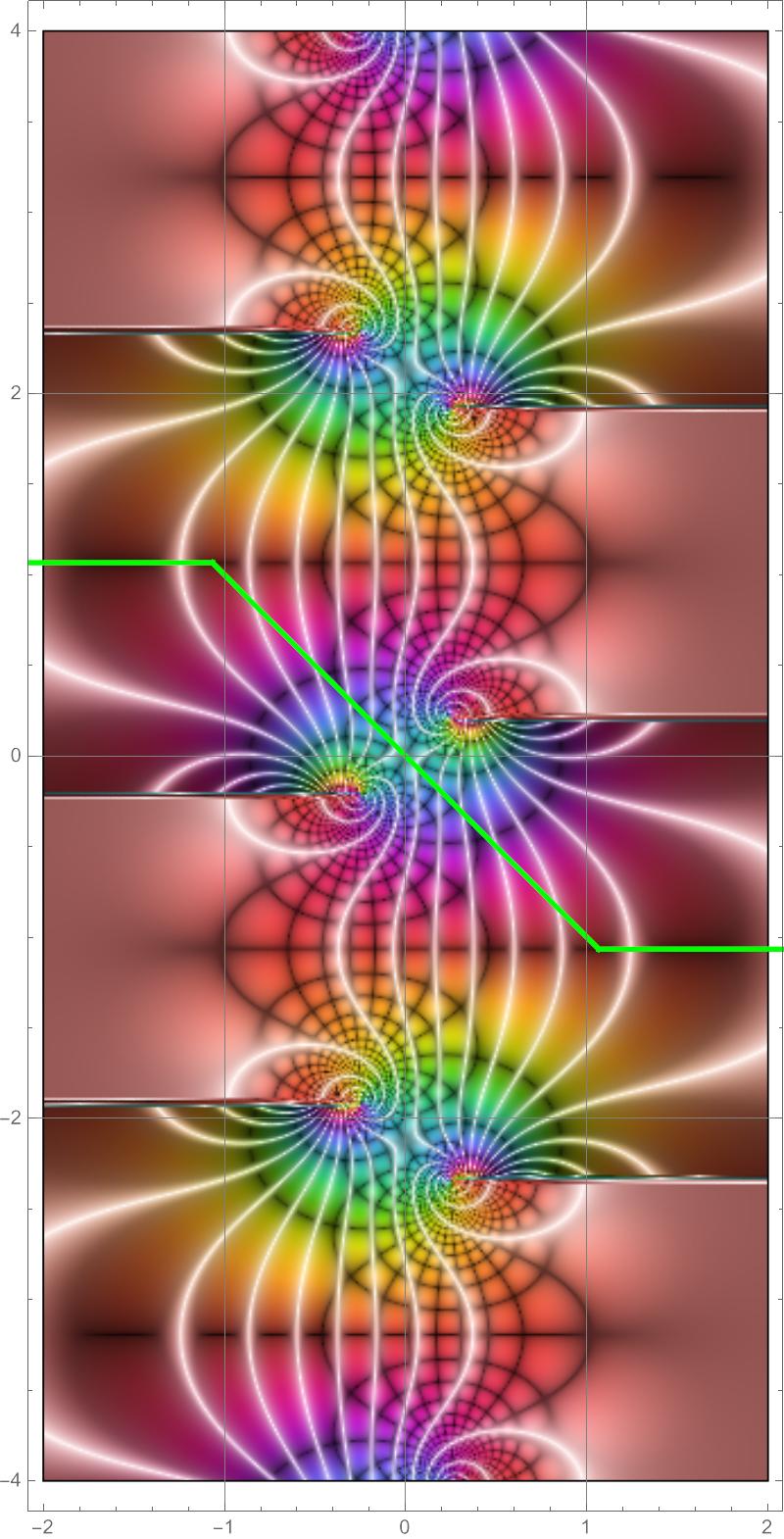}
    \caption{$\beta=0$. $t(u)$, $x(u)$, $\dot{t}(u)$ and $\dot{x}(u)$ in the complex proper-time plane, $-2<\text{Re }u<2$, $-4<\text{Im }u<4$, obtained by solving~\eqref{inflationEqs} with initial conditions $x(0)=\dot{t}(0)=0$ and $t(0)=t_B=\eqref{ASautertB}$ for $\alpha=1$ and $p=p_s$. The results are produced as described in~\cite{DegliEsposti:2023qqu} and plotted with Mathematica's ``ComplexPlot[...,ColorFunction$\to$"CyclicReImLogAbs"]''. Red means real and positive, and blue means real and negative. The black lines are lines where either the real or imaginary part is constant. The bright green line shows one allowed and suitable contour. The corresponding contour is shown in Fig.~\ref{fig:rPlotsBeta0}.}
    \label{fig:complexPlotsBeta0}    
\end{figure*}

\begin{figure*}
    \centering
    \includegraphics[width=0.329\linewidth]{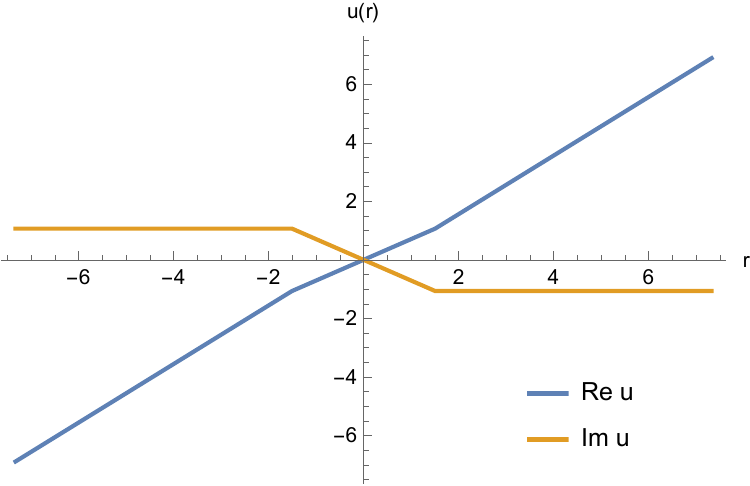}
    \includegraphics[width=0.329\linewidth]{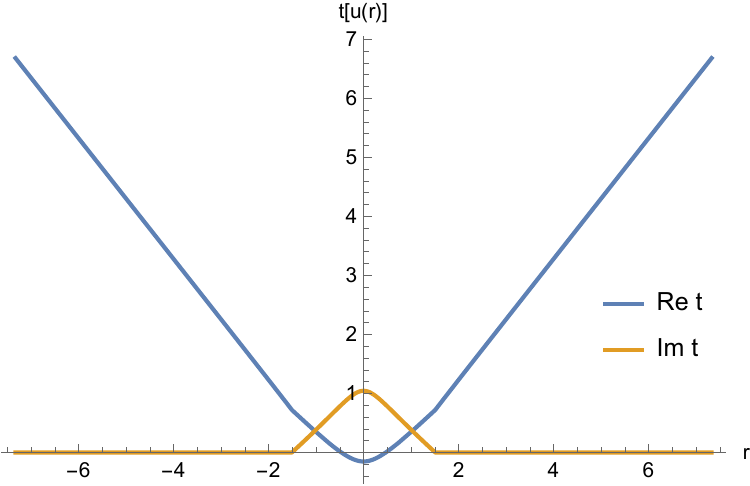}
    \includegraphics[width=0.329\linewidth]{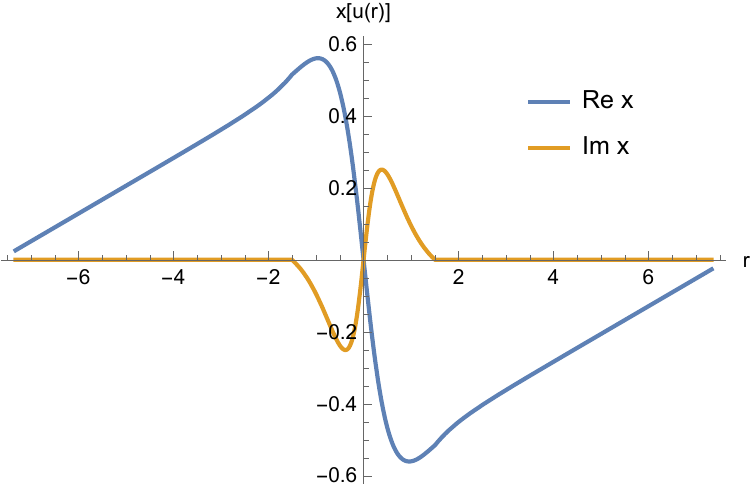}
    \caption{The real and imaginary parts of $u(r)$, $t(r)$ and $x(r)$ for the bright green contour in Fig.~\ref{fig:complexPlotsBeta0}. Note that $r$ is not a radial coordinate, but a real parametrization of the complex proper time contour, $u(r)$.}
    \label{fig:rPlotsBeta0}
\end{figure*}

\section{Metrics on ``Inflation'' form}\label{inflation section}

We begin with metrics in the form
\be\label{inflationType}
\ud s^2=\ud t^2-b(t,x)\ud x^2=\ud t^2-e^{2 A(t,x)}\ud x^2 \;.
\ee
Since de Sitter can be expressed as~\eqref{inflationType} with $A\propto t$, we will refer to~\eqref{inflationType} as ``inflation'' metrics, but we consider~\eqref{inflationType} mainly to illustrate the methods, rather than for a realistic study of inflation. The EOMs are given by
\be\label{inflationEqs}
\begin{split}
\ddot{t}&=-\frac{\dot{x}^2}{2}\partial_t b=(1-\dot{t}^2)\partial_t A\\
\ddot{x}&=-2\dot{t}\dot{x}\partial_t A-\dot{x}^2\partial_x A \;,
\end{split}
\ee
where we used $g_{\mu\nu}\dot{q}^\mu\dot{q}^\nu=1$ to simplify the first line. The $\ddot{x}$ equation can be replaced by
\be\label{inflationDx}
\dot{x}=\sqrt{\dot{t}^2-1}e^{-A} \;,
\ee
which, though, requires some extra work to make the square root continuous along the entire worldline. 

As an example, we consider
\be\label{ASauterSauter}
A=\alpha \tanh(\omega t)\text{sech}^2(kx) \;,
\ee
but our methods work for general $A(t,x)$.
By rescaling $q^\mu\to q^\mu/\omega$ and $u\to u/\omega$, we see that $\mathcal{A}(\omega,k)=\mathcal{A}(\beta)/\omega$ where $\beta=k/\omega$, so the result only depends nontrivially on the ratio of $\omega$ and $k$. 

\subsection{Starting point}

To find instantons we start at $\beta=0$, where we can integrate~\eqref{inflationEqs} once,
\be
\dot{t}=\sqrt{1+p^2e^{-2A(t)}}
\qquad
\dot{x}=pe^{-2A(t)} \;.
\ee
With $A$ normalized as in~\eqref{ASauterSauter}, the asymptotic momentum is given by $\dot{x}(\pm\infty)=pe^{-2\alpha}$. After changing integration variable in~\eqref{generalExpFin} from $u$ to $t$, and performing partial integration, we find
\be
\mathcal{A}=\frac{2}{\omega}\text{Im}\int_\infty^\infty\ud t\sqrt{1+p^2e^{-2A(t)}} \;.
\ee
The integration contour goes around one of the branch points. For~\eqref{ASauterSauter} we have
\be\label{ASautertB}
t_B=\text{arctanh}\left(\frac{1}{2\alpha}\left[i\pi+\ln p^2\right]\right) \;.
\ee
Since the integrand is real for real $t$, we can choose a contour parameterized as $t(s)=\text{Re }t_B+si\text{Im }t_B$, where $0<s<1$. Because of the branch, we have $\int_0^1\ud s\dots+\int_1^0\ud s\dots=2\int_0^1\ud s\dots$. The results are shown in Fig.~\ref{fig:ASauterTime}. For $\alpha\ll1$ we find convergence to the perturbative result, $\mathcal{A}=2\pi\sqrt{1+p^2}/\omega$, and for $\alpha\gg1$ to the de-Sitter result, $\mathcal{A}=2\pi/\omega\alpha$. We will discuss these limits in detail below.

\subsection{Numerical continuation}

\begin{figure*}
    \centering
    \includegraphics[width=.329\linewidth]{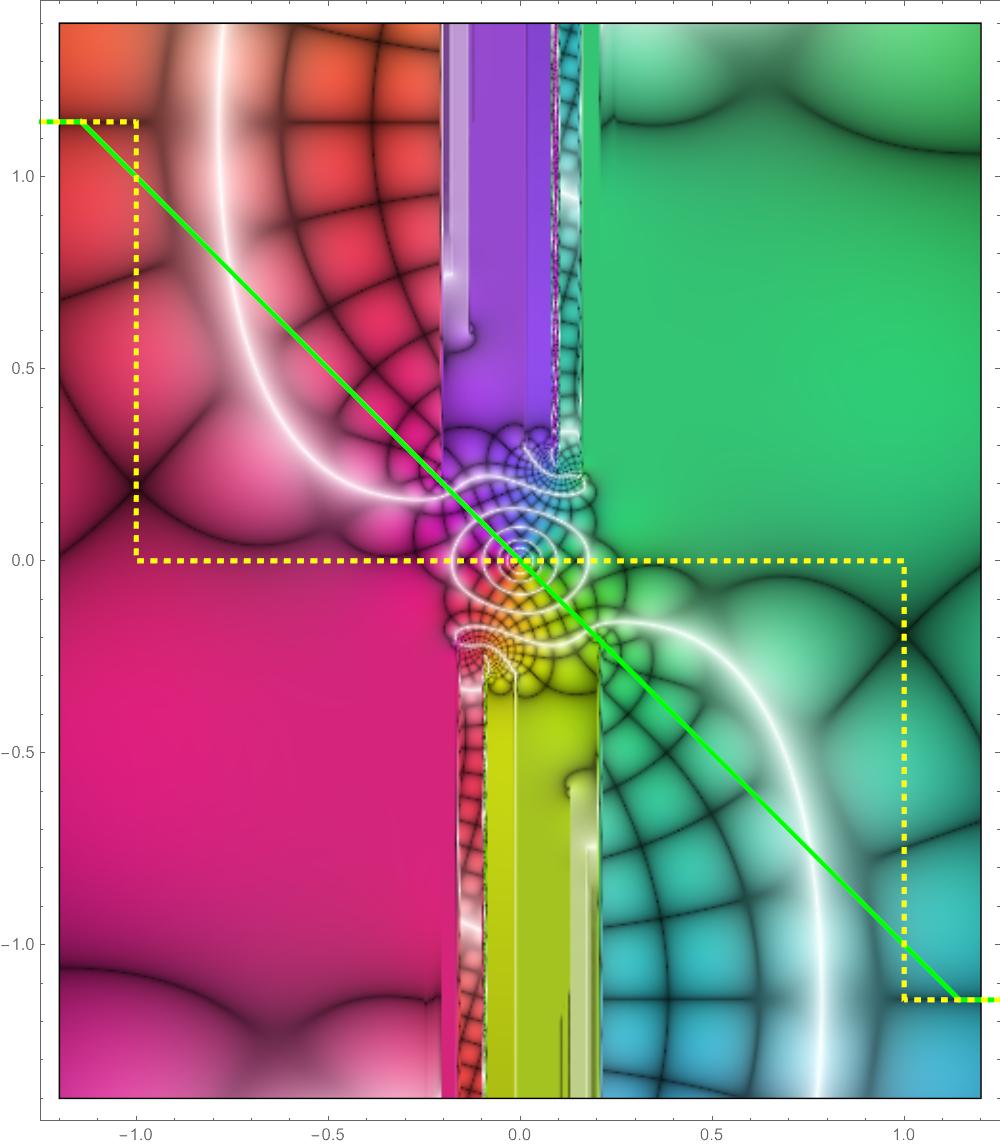}
    \includegraphics[width=.329\linewidth]{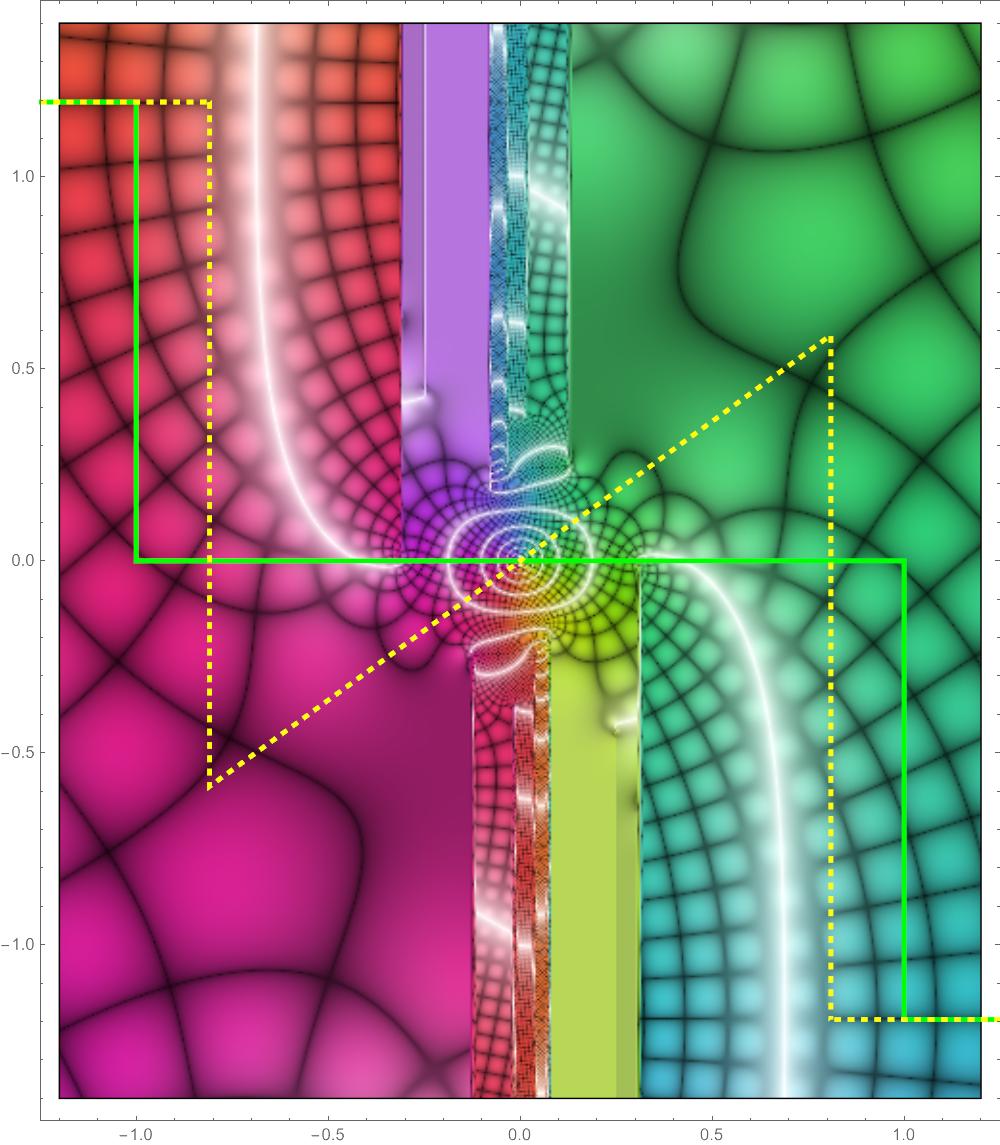}
    \includegraphics[width=.329\linewidth]{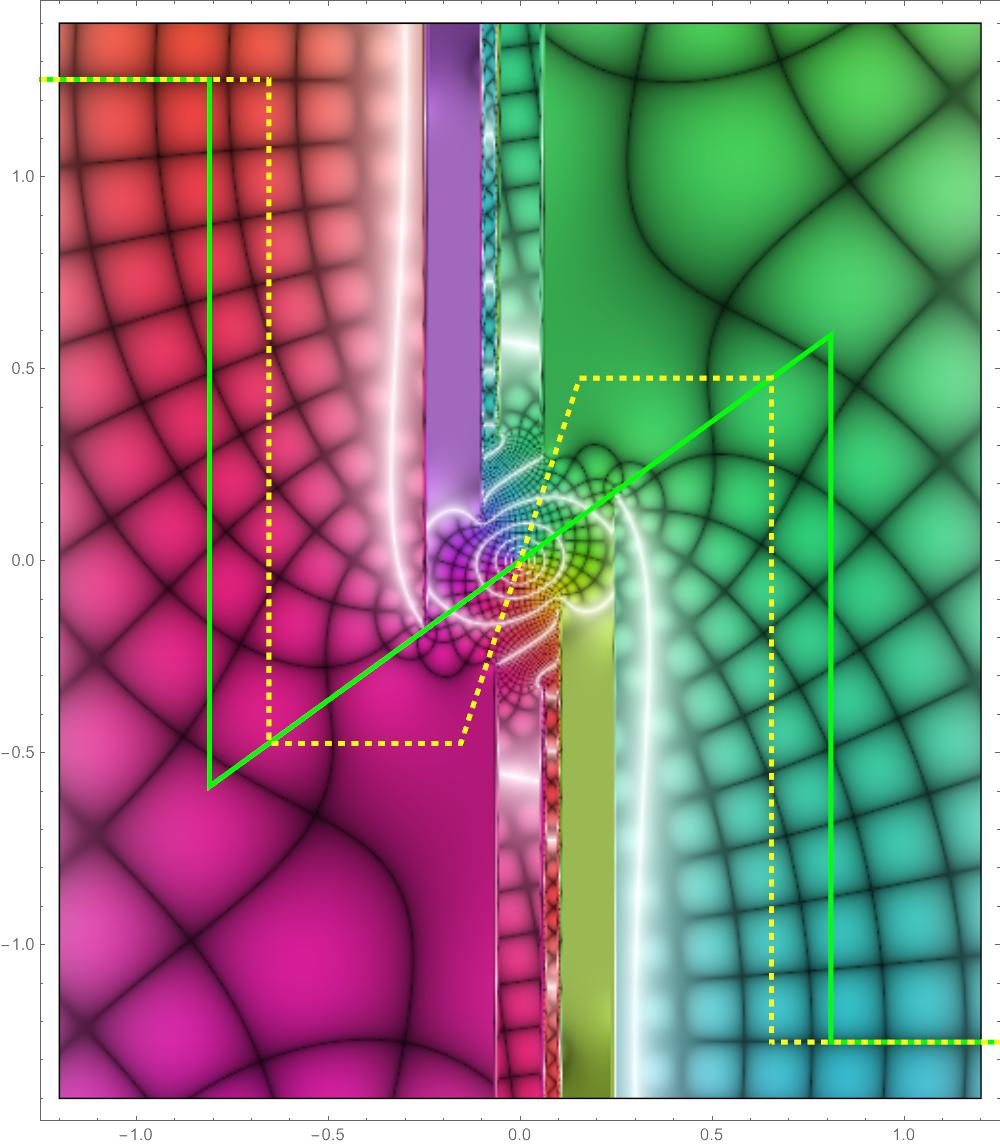}
    \caption{$x(u)$ as in Fig.~\ref{fig:complexPlotsBeta0} but for $\beta=\{3,8,21.8\}$. In each plot, we switch from the solid to the dashed contour to avoid the moving branch points as we numerically continue towards increasing $\beta$.}
    \label{fig:complexPlotBetas}
\end{figure*}

\begin{figure}
    \centering
    \includegraphics[width=.8\linewidth]{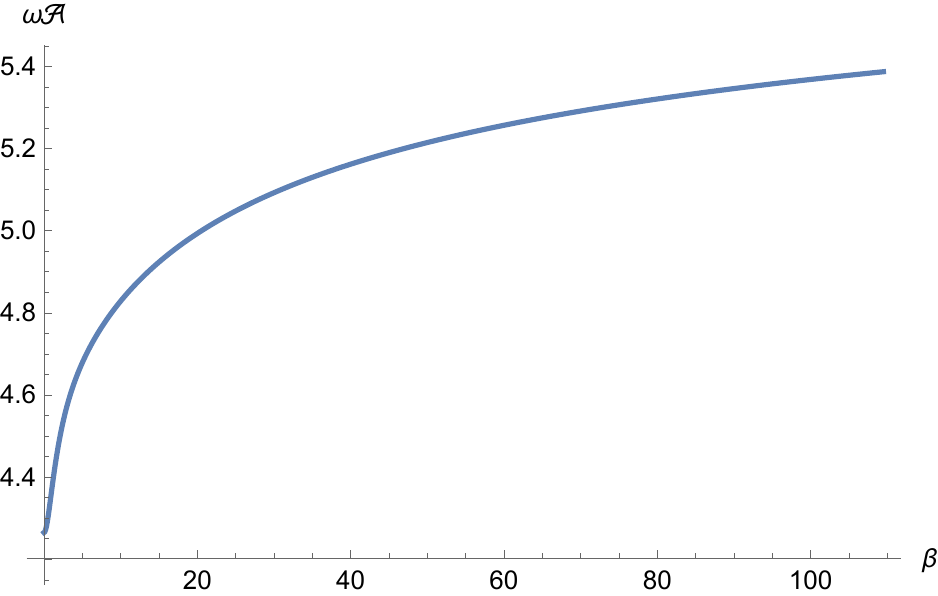}
    \caption{$\mathcal{A}(\beta)$ for~\eqref{ASauterSauter} with $\alpha=1$ and for the saddle-point values of the momenta.}
    \label{fig:AofBeta}
\end{figure}

\begin{figure*}
    \centering
    \includegraphics[width=.329\linewidth]{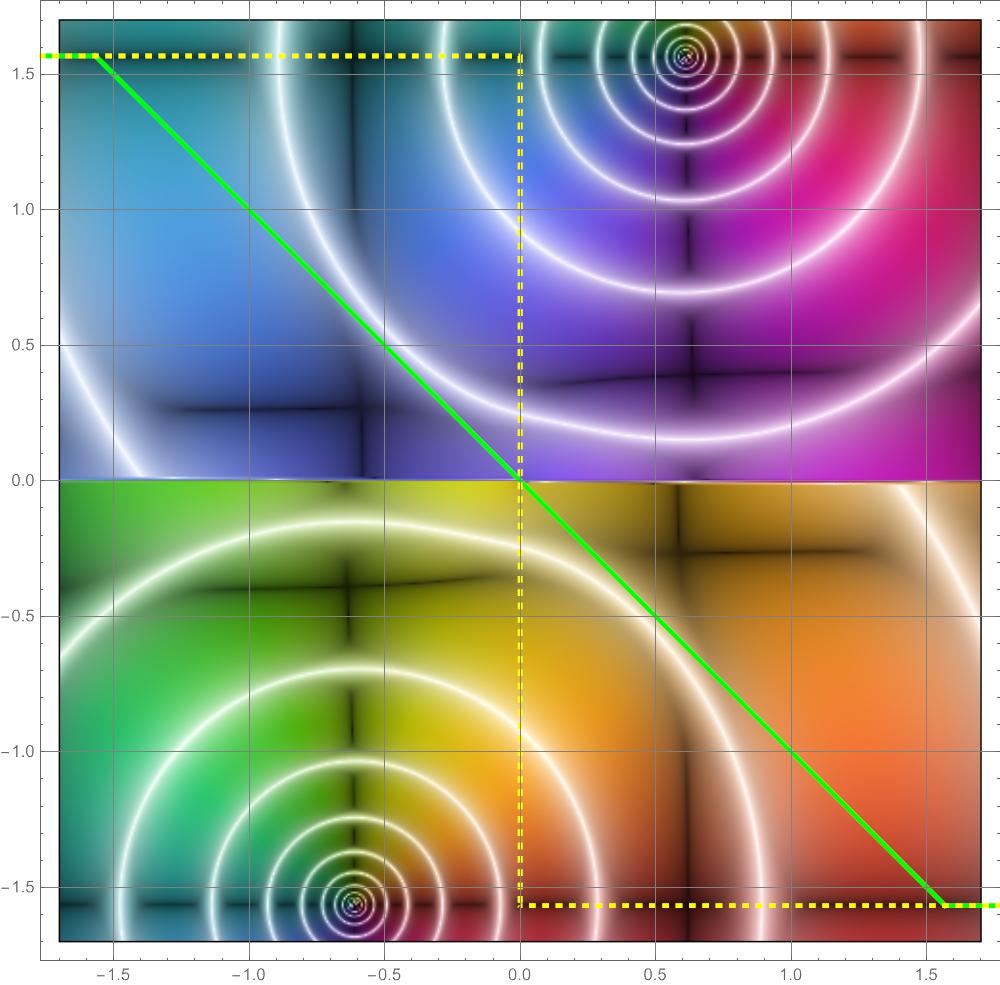}
    \includegraphics[width=.329\linewidth]{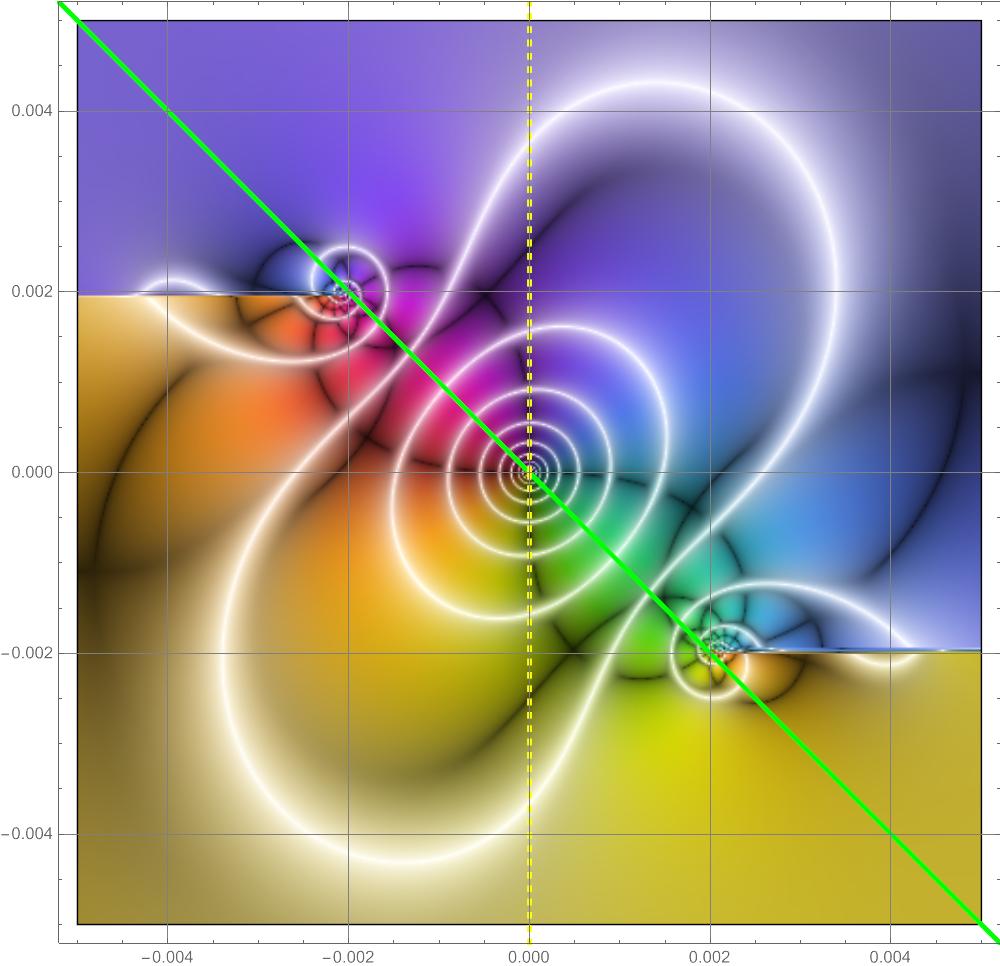}
    \includegraphics[width=.329\linewidth]{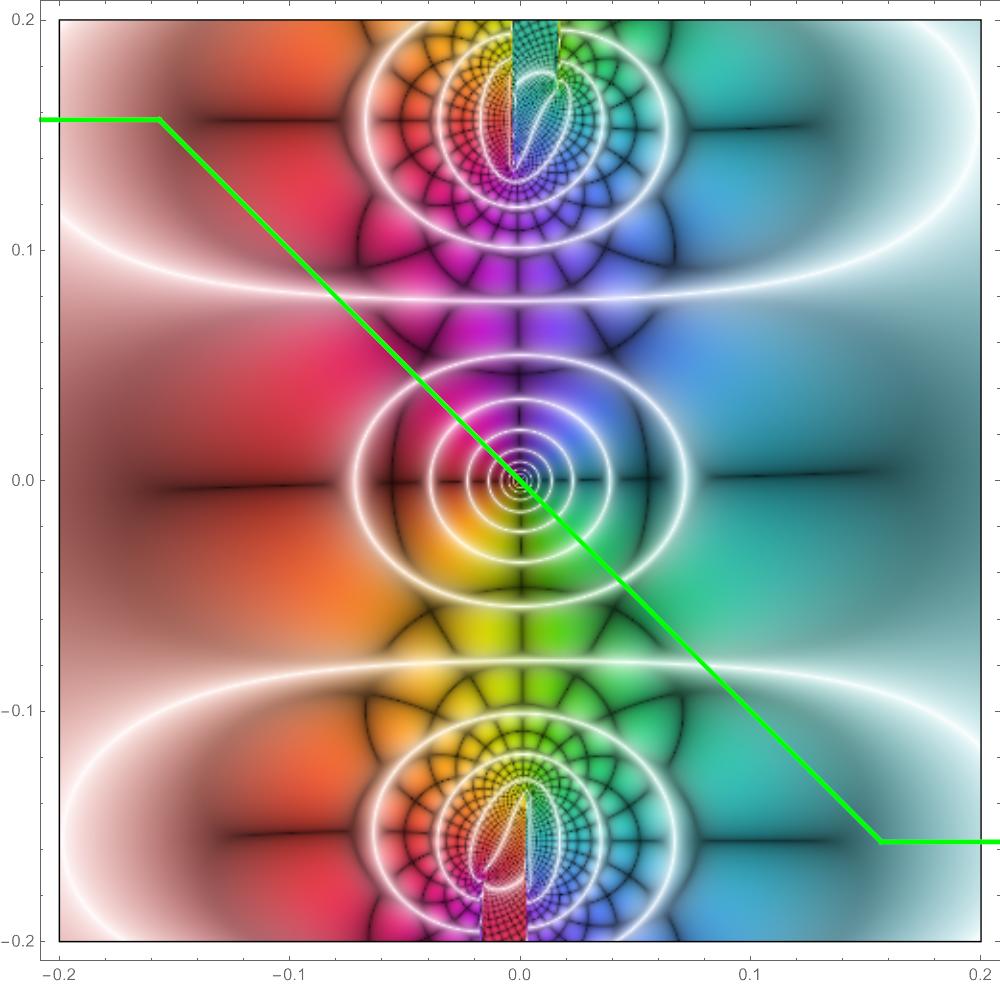}
    \caption{$x(u)$ as in~\ref{fig:complexPlotsBeta0} but for $\beta=1$, $\alpha=0.03$ in the first two plots and $\alpha=10$ in the third. The second plot zooms in on a small region in the first plot around $u=0$.}
    \label{fig:complexPlotBeta1alphas}
\end{figure*}

We could obtain the $\beta=0$ results in Fig.~\ref{fig:ASauterTime} without having to find explicit expressions for the instantons, but we cannot do that for $\beta>0$. To be able to use the $\beta=0$ limit as a starting point for $\beta>0$, we have to find the instantons. As an example, we take the result in Fig.~\ref{fig:ASauterTime} for $\alpha=1$ and $p=p_s\approx0.57$. Without loss of generality, we can choose initial conditions for~\eqref{inflationEqs} such that $\dot{t}(u=0)=0$. The value of $x(0)$ is irrelevant for $\beta=0$, but for $\beta>0$ we have $x(0)=0$ for the instanton that maximizes the probability. For $\beta=0$ we have $t(0)=t_B$, where $t_B\approx-0.15+1.04i$ is given by~\eqref{ASautertB}. $\dot{x}(0)$ is then given by~\eqref{inflationDx}. 

Now that we have figured out $q^\mu(0)$ and $\dot{q}^\mu(0)$, we need to find a suitable $u$ contour. We start by making plots of $t(u)$ and $x(u)$ in the complex $u$ plane. This is done, as in~\cite{DegliEsposti:2023qqu}, by solving~\eqref{inflationEqs} on a dense set of lines, making an $M\times N$ table of $t(u_{i,j})$ etc., with $1<i<M$ and $1<j<N$, making an interpolation function of the table, and then plotting the result. The results are shown in Fig.~\ref{fig:complexPlotsBeta0}. Producing such plots for $M,N\sim 200$ only takes a couple of seconds with Mathematica on a standard laptop, which is fortunate since we need to make such plots each time we need to figure out a suitable $u$ contour, and it turns out that we need to do that quite often, which is not something that one would have expected based on the QED cases in~\cite{DegliEsposti:2023qqu,DegliEsposti:2022yqw,DegliEsposti:2024upq}. 

It is in general not possible to find contours such that $t$ and $x$ are both real asymptotically. However, in this example we focus on the saddle-point values of the momenta, and then we can find contours such that all components of $q^\mu(\pm\infty)$ and $\dot{q}^\mu(\pm\infty)$ are real. Since both particles should travel towards $t\to+\infty$, we need $\dot{t}(+\infty)>1$ and $\dot{t}(-\infty)<-1$ (or vice versa). The goal is to find $q^\mu(0)$ and $\dot{q}^\mu(0)$ such that $q$ satisfies~\eqref{saddleCondition}. From Fig.~\ref{fig:complexPlotsBeta0}, we see that in this case we could let $u$ follow the real axis. However, that is not the most efficient contour, because we would need to integrate out to larger values of $|u|$ to check~\eqref{saddleCondition}. If we instead choose the green contour in Fig.~\ref{fig:complexPlotsBeta0}, then we can stop integrating at the two points where $q^\mu(u)$ and $\dot{q}^\mu(u)$ become real (where the diagonal line becomes horizontal), because then $q^\mu(u)$ and $\dot{q}^\mu(u)$ will stay real on a line parallel with $\text{Im }u=\text{const.}$, which means both sides of~\eqref{saddleCondition} will stay zero. The goal then is to find $q(0)$ and $\dot{q}(0)$ such that $\text{Im }q(u_c)=\text{Im }\dot{q}(u_c)=0$ at some $u_c$. We have already done that for $\beta=0$, as seen in Fig.~\ref{fig:complexPlotsBeta0}.  

We will now numerically continue from $\beta=0$ to, say, $\beta=100$. Initially we choose a step size of $\Delta\beta=0.1$.  
For $0<\beta\ll1$ we can still use the same $u$ contour as in Fig.~\ref{fig:complexPlotsBeta0}. We still have $x(0)=0$ due to symmetry, so we only need to find two real constants, $\text{Re }t(0)$ and $\text{Im }t(0)$. We therefore only need to impose two of the four conditions in~\eqref{saddleCondition}, e.g. the two at $u=u_1$. Thus, we start at $u=0$ and integrate down along the diagonal line until $\text{Im }t[u(r_s)]=0$, where the code stops. We repeat with different values of $t(0)$ until $\text{Im }x[u(r_s)]=0$ and $\text{Im }\dot{t}[u(r_s)]=0$ or $\text{Im }\dot{x}[u(r_s)]=0$. If $\text{Im }\dot{t}[u(r_s)]=0$ then $\text{Im }\dot{x}[u(r_s)]=0$ automatically, and vice versa, so we only have to check one of them. For this Newton-Raphson method to converge, we need a good starting point. For the first step, $\beta=\beta_1=\Delta\beta\ll1$, we can use the result for $\beta=0$ as starting point, because $t(0,\Delta\beta)\approx t(0,\beta=0)=t_B=\eqref{ASautertB}$. For $\beta_n=n\Delta\beta$ we can use the result for $\beta_{n-1}$ as starting point, but for $n>3$ we can significantly speed up this process by making a quadratic extrapolation,
\be
\begin{split}
Y(\beta_n)&\approx\frac{(\beta_{n}-\beta_{n-1})(\beta_{n}-\beta_{n-2})}{(\beta_{n-3}-\beta_{n-1})(\beta_{n-3}-\beta_{n-2})}Y(\beta_{n-3})\\
&+\frac{(\beta_{n}-\beta_{n-1})(\beta_{n}-\beta_{n-3})}{(\beta_{n-2}-\beta_{n-1})(\beta_{n-2}-\beta_{n-3})}Y(\beta_{n-2})\\
&+\frac{(\beta_{n}-\beta_{n-2})(\beta_{n}-\beta_{n-3})}{(\beta_{n-1}-\beta_{n-2})(\beta_{n-1}-\beta_{n-3})}Y(\beta_{n-1}) \;,
\end{split}
\ee
where $Y(\beta)=t(0,\beta)$ in this case. 

Using the above algorithm, we can quickly obtain $\mathcal{A}(\beta)$ up to $\beta=3$, but at that point the code breaks. To figure out why, we make new complex plots. By comparing Figs.~\ref{fig:complexPlotsBeta0} and the first plot in~\ref{fig:complexPlotBetas} we see that, as we increased $\beta$ from $0$ to $3$, the branch points moved towards the $u$ contour. To continue past $\beta>3$ we therefore need to choose a new contour. We choose the dashed curve. That contour works until $\beta=8$, where we again need to choose a new contour, as shown in the second plot in Fig.~\ref{fig:complexPlotBetas}. At $\beta=21.8$ we again need a new contour, as shown in the third plot in Fig.~\ref{fig:complexPlotBetas}. Now we can continue up to $\beta=100$ without having to choose a new contour. Fig.~\ref{fig:AofBeta} shows that $\mathcal{A}(\beta)$ is continuous and smooth, including at the points where we changed the contour.    

The above procedure is representative for the metrics we have considered. One often has to be careful when navigating around the branch points, which move around as one changes the values of the metric parameters or the momenta. From the complex plots we see that there are multiple branch points on the plotted part of the Riemann surface. There are additional branch points hidden on other Riemann sheets. If we consider a contour which is not continuously deformable to the ones in Fig.~\ref{fig:complexPlotBetas}, then we should in general expect to find a non-equivalent instanton, even for the same $q^\mu(0)$ and $\dot{q}^\mu(0)$. The existence of multiple branch points then suggests that there could be a large number of potentially relevant non-equivalent contours. Some of them can be ruled out immediately since they give non-physical results, e.g. $\mathcal{A}<0$. One way to check that one has considered the relevant contour(s) is to check various limits. We have in some sense already checked the limit where we started the numerical continuation, $\beta=0$ in the above example. Next we will study the $\alpha\ll1$ and $\alpha\gg1$ limits.      

\subsection{Perturbative limit}

For $\alpha\ll1$ one can expect to be able to treat the pair-production probability using perturbation theory in $h_{\mu\nu}=g_{\mu\nu}-\eta_{\mu\nu}$ instead of the instanton approach. The probability for general metrics to $\mathcal{O}(h^2)$ can be found in~\cite{Frieman:1985fr,Boasso:2024ryt}. The exponential part of the probability is determined by the Fourier transform of the metric, which for a time-dependent metric is given by (omitting the indices)
\be
h(w)=\int\ud t\, h(t)e^{iwt} \;.
\ee
For fields like~\eqref{ASauterSauter} or $1/(1+[\omega t]^2)$ which have a pole at $t=i\nu/\omega$, for some constant $\nu$, $h(w)$ has an exponential scaling for the relevant values of $w$,
\be\label{fourierExp}
h(w)\sim\exp\left(-\nu\frac{w}{\omega}\right) \;.
\ee
To produce a pair with energy $2p_0$ by the absorption of $n$ ``gravitons'', we need (assuming symmetry) $w=2p_0/n$ from each graviton, so the contribution to the probability scales as
\be\label{perturbativeFourier}
P_n\sim |h^n(w)|^2\sim\exp\left(-\frac{4\nu}{\omega}p_0\right) \;.
\ee
Since the exponential in~\eqref{perturbativeFourier} is the same for each order, one can expect that a resummation of $\sum_n P_n$ should have the same exponential. Thus, we expect the weak-field limit of our instanton result should converge to the perturbative result,
\be\label{AperturbativeLimit}
\lim_{\alpha\to0}\mathcal{A}=-\frac{4\nu}{\omega}p_0 \;.
\ee
For other types of fields, e.g. $e^{-(\omega t)^2}$, there is in general a dominant order, $n_{\rm dom}$, that can be large~\cite{Torgrimsson:2017pzs}. One can still expect $\alpha\ll1$ to be a perturbative limit, but there would not be any simple relation such as~\eqref{AperturbativeLimit}, so we would not be able to check $\lim_{\alpha\ll1}\mathcal{A}$. Since it is quite useful and reassuring to be able to check the $\alpha\ll1$ limit of our generally nonperturbative results by comparing with a completely independent perturbative calculation, we prefer to consider fields of the~\eqref{fourierExp} type.   

\begin{figure}
    \centering
    \includegraphics[width=\linewidth]{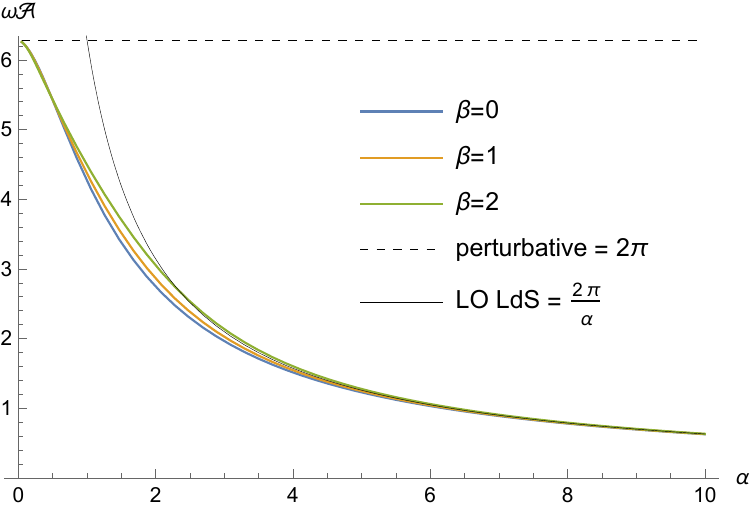}
    \caption{$\mathcal{A}$ as a function of $\alpha$ for $\beta=k/\omega=0,1,2$ compared to the perturbative result (\eqref{perturbativeFourier} with $p=0$) and the LO LdS approximation (first term in~\eqref{A1and3}), for the metric in~\eqref{inflationMetric} with $f'(u)=F(u)=\text{sech}^2u$.}
    \label{fig:AofAlpha}
\end{figure}

\begin{figure}
    \centering
    \includegraphics[width=\linewidth]{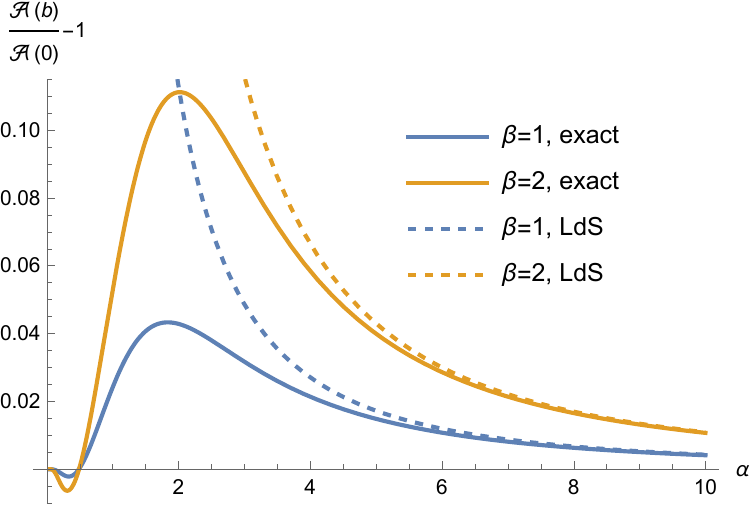}
    \caption{Same as in Fig.~\ref{fig:AofAlpha}, showing the relative difference for $\beta\ne0$ and $\beta=0$. The dashed lines are obtained by including both the LO and the NLO terms in~\eqref{A1and3}. Since the LO does not depend on $\beta$, the result would be zero without the NLO.}
    \label{fig:AbA0}
\end{figure}

\subsection{Locally de-Sitter approximation}

We will now derive a LdS approximation for the opposite limit, $\alpha\gg1$. We assume that the metric can be expressed as
\be\label{inflationMetric}
\ud s^2=\ud t^2-\exp\left\{2\alpha f(\omega t)F(kx)\right\}\ud x^2 \;,
\ee
where $f(-u)=-f(u)$ is a monotonically increasing function normalized such that $f'(0)=1$, and $F(-u)=F(u)$ with $F(0)=1$.
We begin by rescaling $q^\mu\to q^\mu/\alpha$ and $u\to u/\alpha$. 
We can expand the instanton as $q^\mu=q_{(0)}^\mu+\alpha^{-2}q_{(2)}^\mu+\mathcal{O}(\alpha^{-4})$ and the exponent as $\mathcal{A}\approx\mathcal{A}_1+\mathcal{A}_3$. The solution of~\eqref{inflationEqs} is to LO
\be
t_{(0)}(u)=\frac{i\pi}{2}+\ln[p\cosh(u)]
\qquad
x_{(0)}(u)=-\frac{\tanh(u)}{p}\;,
\ee
where $p$ is a constant (for $k=0$ it has a simple relation to the asymptotic momentum). Plugging this into~\eqref{generalExpFin} gives $\mathcal{A}_1=2\pi/(\alpha\omega)$ independently of $p$. This agrees with the flat line for $\alpha=8$ in Fig.~\ref{fig:ASauterTime}. Thus, the leading order agrees with what one would expect for pair production in de Sitter~\cite{Mottola:1984ar}.
As noted already in~\cite{DegliEsposti:2023qqu,DegliEsposti:2023fbv}, since we evaluate all the integration variables at their saddle-point values, one can obtain $\mathcal{A}_3$ using only $q_{(0)}^\mu$ without the need to calculate $q_{(2)}^\mu$ by expanding the original version of the exponent~\eqref{worldlineAction} treating all the integration variables as independent of $\alpha$. We find
\be
\mathcal{A}_3=\left(-\frac{\pi}{6}[\pi^2-6]+\pi\ln^2\left[\frac{ep}{2}\right]\right)\kappa_0+\frac{\pi}{3p^2}\kappa_1 \;,
\ee
where $\kappa_0=-f'''(0)>0$ and $\kappa_1=-(k/\omega)^2F''(0)>0$. For~\eqref{ASauterSauter}, $f'(u)=F(u)=\text{sech}^2u$, we have $\kappa_0=\kappa_1=2$. For a purely time dependent metric, $\kappa_1=0$, we find that $\mathcal{A}_3$ is minimized by $p=2/e\approx0.74$, which is consistent with the limit suggested by Fig.~\ref{fig:ASauterTime}. For time and space dependent metrics we find instead a saddle point at
\be
p=\sqrt{\frac{2\kappa_1}{3\kappa_0}}\left[W\left(\frac{e^2\kappa_1}{6\kappa_0}\right)\right]^{-1/2} \;,
\ee
where $W(z)$ is the Lambert W function ($z=W e^W$). Plugging this into $\mathcal{A}_3$ we find the LdS approximation up to next-to-leading order (NLO),
\be\label{A1and3}
\begin{split}
\mathcal{A}&\approx\frac{2\pi}{\alpha\omega}\\
&+\frac{\pi\kappa_0}{6\alpha^3\omega}\left\{6-\pi^2+\frac{3}{2}W\left[\frac{e^2\kappa_1}{6\kappa_0}\right]\left(2+W\left[\frac{e^2\kappa_1}{6\kappa_0}\right]\right)\right\} \;.
\end{split}
\ee
$\mathcal{A}$ is a monotonically increasing function of $k$, so making the metric narrower in the $x$ direction makes the probability more suppressed. This trend can also be seen in the $\alpha=1$ result in Fig.~\ref{fig:AofBeta}, and is in general also what one would expect in the QED case (see e.g.~\cite{Dunne:2006st,Ilderton:2015qda}). 

\subsection{Dependence on $\alpha$}

In obtaining Fig.~\ref{fig:AofBeta}, we have obtained $t(0)$ for $\alpha=1$ and $0<\beta<100$. We can now use that as starting points for a numerical continuation in $\alpha$ for fixed $\beta$. Fig.~\ref{fig:complexPlotBeta1alphas} shows what contours are possible for $\beta=1$ and $0<\alpha<10$. For $\alpha>0.03$ we can use the tilted step contour in Fig.~\ref{fig:complexPlotsBeta0}, while for smaller $\alpha$ we can use the dashed contour in Fig.~\ref{fig:complexPlotBeta1alphas} instead.  

Fig.~\eqref{fig:AofAlpha} shows the results for a couple of different values of $\beta$. For $\alpha\ll1$ we see convergence to the perturbative result~\eqref{perturbativeFourier} for the saddle-point value of the momentum, $p=0$. For $\alpha\gg1$ we find convergence to the LdS approximation~\eqref{A1and3}. From Fig.~\eqref{fig:AofAlpha} it might at first seem like the effect of the spatial dependence is quite modest, but Fig.~\ref{fig:AbA0} shows that the relative difference $[\mathcal{A}(b=2)/\mathcal{A}(b=0)]-1$ is at least more than $10\%$, and, since $\mathcal{A}\propto1/\omega\gg1$, the corresponding difference in the probability can be quite large, $P\sim e^{-\mathcal{A}(b)}\ll e^{-\mathcal{A}(0)}$. Fig.~\ref{fig:AbA0} also demonstrates the correctness of the NLO term in the LdS approximation~\eqref{A1and3}. 

\begin{figure*}
    \centering
    \includegraphics[width=0.329\linewidth]{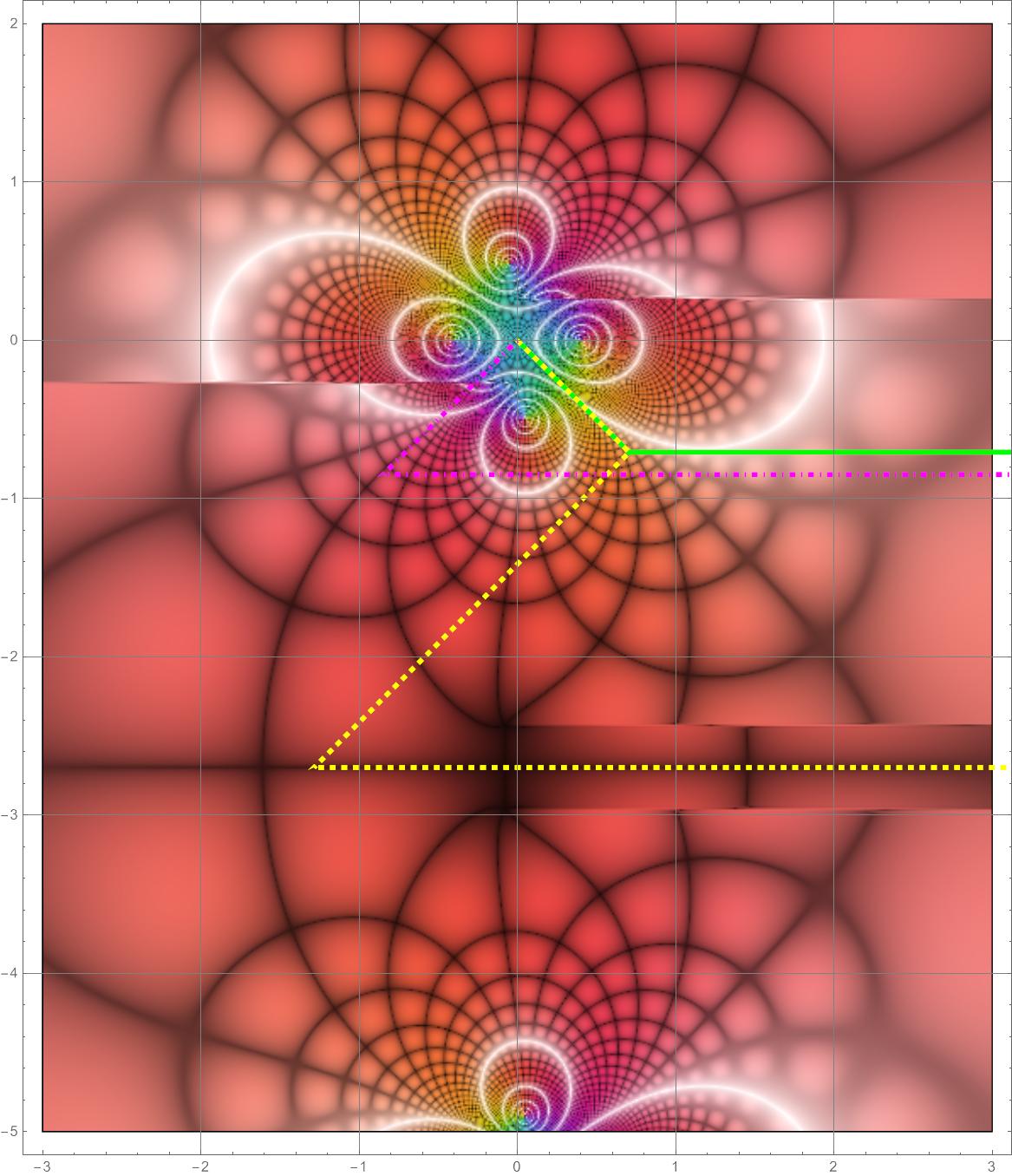}
    \includegraphics[width=0.329\linewidth]{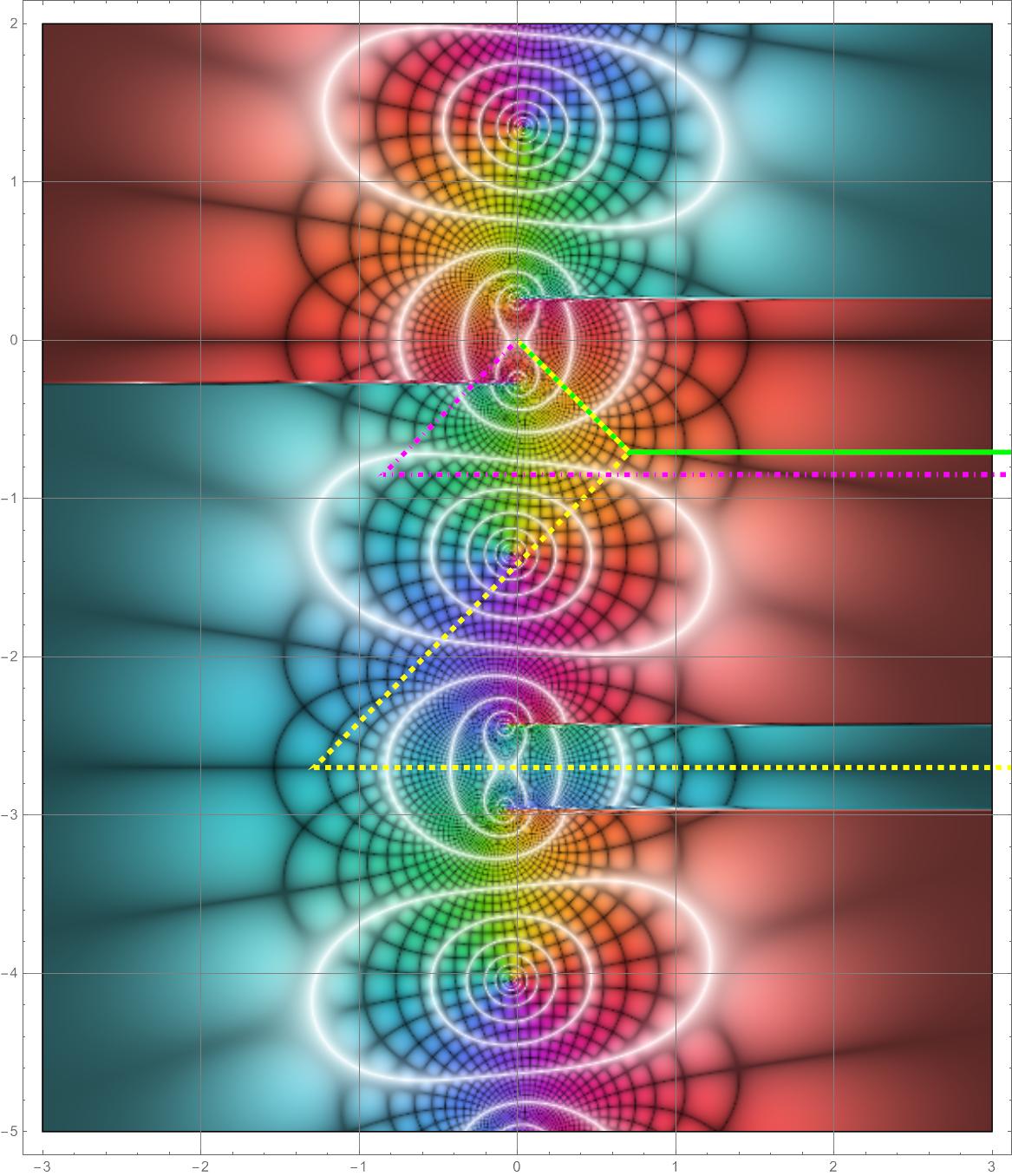}
    \includegraphics[width=0.329\linewidth]{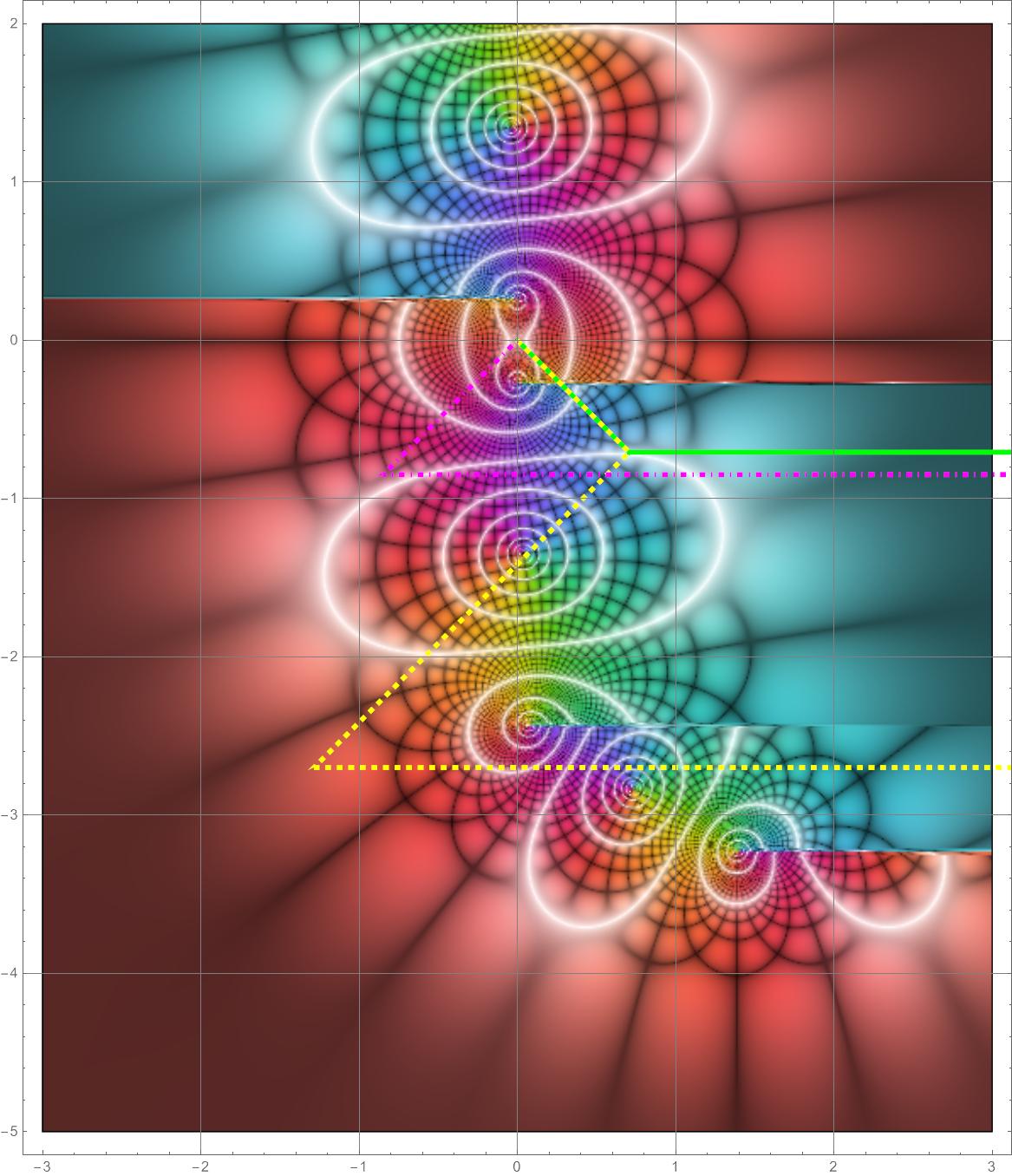}
    \caption{$\dot{t}(u)$ (first plot) and $\dot{x}(u)$ (second and third plots) for~\eqref{vPulse} with $\alpha=2$, $k=1$, $\omega=0$ and $p^1=0.5$. The two $\dot{x}(u)$ plots show different parts of the same Riemann surface. The solid, dashed and dot-dashed lines show one ``half'' of the contour. The other half goes from $u=0$ to $u=u_I\approx-0.4$, where $\dot{t}$ has a pole due to the inner horizon.}
    \label{fig:xComplexPlot3start}
\end{figure*}

\begin{figure*}
    \centering
    \includegraphics[width=0.4\linewidth]{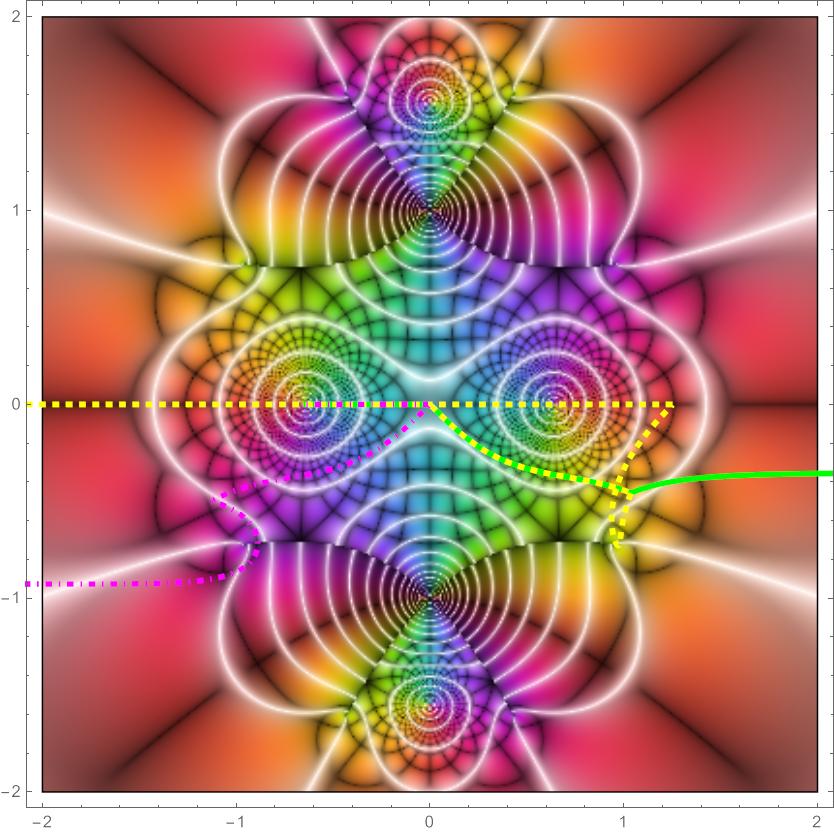}
    \includegraphics[width=0.4\linewidth]{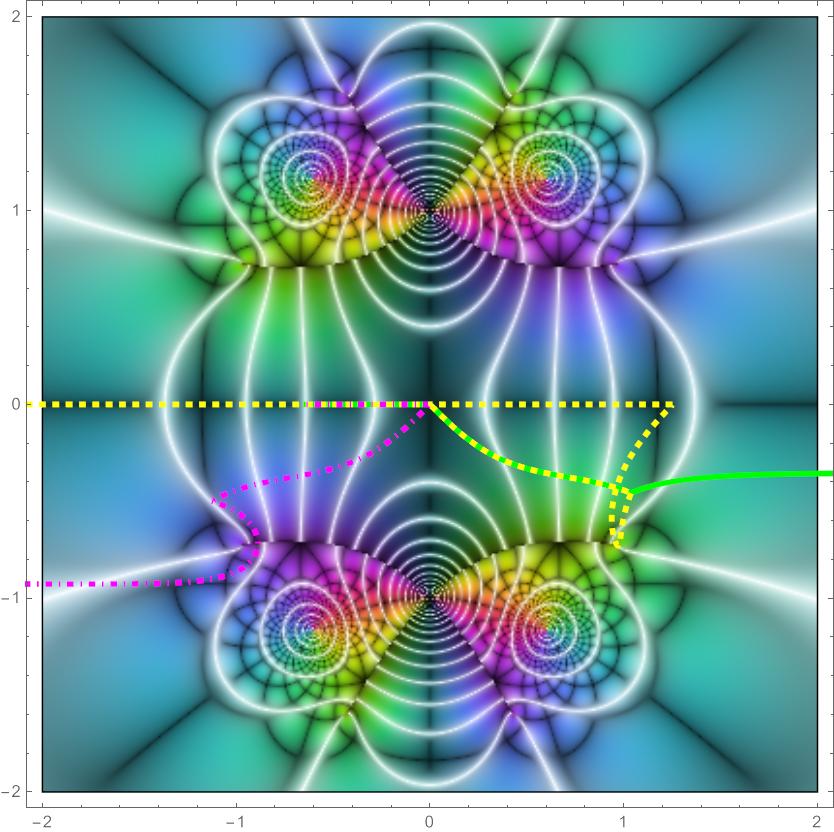}
    \caption{The integrand in~\eqref{AGPstat} in the complex $x$ plane. The two plots show the two parts of the Riemann surface. The three contours show $x(u)$ for the three $u$ contours in Fig.~\ref{fig:xComplexPlot3start}.}
    \label{fig:Agrand3start}
\end{figure*}

\section{Metrics on Gullstrand-Painlev\'e form}\label{acoustic section}

As a second class of coordinates, we consider Gullstrand-Painlev\'e (GP) coordinates,
\be\label{ds2GP}
\ud s^2=[1-v^2(t,x)]\ud t^2+2v(t,x)\ud t\ud x-\ud x^2 \;.
\ee
See~\cite{Visser:2001kq,Nielsen:2005af} for discussions of~\eqref{ds2GP} and the more general case where $[1-v^2(t,x)]\ud t^2\to[c^2(t,x)-v^2(t,x)]\ud t^2$. Exact solutions for $c(x)$ and $v=\text{const.}$ have been found in~\cite{Fabbri:2015vpt,Anderson:2024mjp}. While~\eqref{inflationType} is convenient for starting a numerical continuation at some $g_{\mu\nu}(t)$, GP is instead convenient for starting at some $g_{\mu\nu}(x)$. $x$ could be a radial coordinate, and for the Schwarzshild metric $v=\sqrt{r_0/r}$. We will use $x$ instead of $r$ for the spatial coordinate. $r$ will instead denote a real parametrization of the complex proper-time contour, $u(r)$ with $-\infty<r<\infty$. GP is often used for analog/acoustic black holes~\cite{Unruh:1980cg,Unruh:1994je,Corley:1998rk,Volovik:1999fc,Schutzhold:2002rf,Unruh:2004zk,Rousseaux:2007is,Macher:2009tw,Macher:2009nz,Rousseaux:2010md,Weinfurtner:2010nu,Mayoral:2010ck,Coutant:2012zh,Balbinot:2019mei,Gaur:2023ved,DelPorro:2024tuw,Barcelo:2005fc,Schutzhold:2025qna}, where $v(x)$ is the background fluid velocity. Our immediate goal in this paper is to use $v(t,x)$ to illustrate our methods.

The geodesic equations are given by
\be\label{GPeqs}
\begin{split}
\ddot{t}&=-(\dot{x}-v\dot{t})^2\partial_x v=(1-\dot{t}^2)\partial_x v\\
\ddot{x}&=[(1-v^2)\dot{t}^2+2v\dot{t}\dot{x}-\dot{x}^2]v\partial_x v+\dot{t}^2\partial_t v\\
&=v\partial_x v+\dot{t}^2\partial_t v \;,
\end{split}
\ee
where the second equal signs are obtained by simplifying~\eqref{geodesicEq} using the on-shell condition~\eqref{TsaddleEq2}. One can replace the $\ddot{x}$ (or the $\ddot{t}$) equation with~\eqref{TsaddleEq2}, which gives 
\be\label{dxFromSqrt}
\dot{x}=v\dot{t}-e^{-i\sigma(u)/2}\sqrt{e^{i\sigma(u)}(\dot{t}^2-1)} \;,
\ee
where $\sqrt{z}$ denotes the square root with a branch cut along, say, $z<0$. $\sigma(u)$ has to be chosen such that~\eqref{dxFromSqrt} is continuous along the entire worldline, which means some extra work compared to using both equations in~\eqref{GPeqs}. But~\eqref{dxFromSqrt} can be useful e.g. for rapidly varying fields.

To find instantons, we use the same ideas as in Sec.~\ref{inflation section}, i.e. we make an initial guess for $q^\mu(0)$ and $\dot{q}^\mu(0)$, and then solve~\eqref{GPeqs} several times using the Newton-Raphson method until we find values of $q^\mu(0)$ and $\dot{q}^\mu(0)$ that give an instanton satisfying the asymptotic conditions~\eqref{asymptoticMomentum2} or~\eqref{saddleCondition}. In Sec.~\ref{inflation section} we used the $g_{\mu\nu}(t,x)\to g_{\mu\nu}(t)$ limit as a starting point. For~\eqref{GPeqs} we will instead start with $v(x)$.
The values of $q^\mu(0)$ and $\dot{q}^\mu(0)$ are not uniquely determined. It is often convenient to choose
\be\label{dt0dxi}
\dot{t}(0)=0 \qquad \dot{x}(0)=\pm i \;.
\ee
If $q(u)$ is a solution to~\eqref{GPeqs} with $\dot{x}(0)=i$, then the solution for $\dot{x}(0)=-i$ is given by $\tilde{q}(u)=q(-u)$, which has the asymptotic momenta in~\eqref{asymptoticMomentum2} swapped $p_\mu\leftrightarrow p_\mu'$. $\tilde{q}$ gives the same $\mathcal{A}$ as $q$.

\subsection{$g_{\mu\nu}(x)$ as starting point for $g_{\mu\nu}(t,x)$}

In this subsection, we consider~\eqref{abc}, where $a,b,c$ only depend on $x$. The $\mu=0$ component of~\eqref{geodesicEq} becomes a total derivative,
$\partial_u[a(x)\dot{t}+c(x)\dot{x}]=0$. This together with~\eqref{TsaddleEq2} give, assuming $a,b\to1$ and $c\to0$ as $x\to\pm\infty$, 
\be\label{dtdxabcStat}
\dot{t}=\frac{1}{a(x)}[p_0-c(x)\dot{x}] 
\qquad
\dot{x}=\epsilon\frac{\sqrt{p_0^2-a}}{\sqrt{ab+c^2}}  \;,
\ee
where $\epsilon=\pm1$.
Since the metric only depends on $x$, we can change variable in~\eqref{generalExpFin} from $u$ to $x$,
\be\label{Aabc}
\mathcal{A}=2\text{Im}\int_{-\infty}^\infty\ud x\, x\partial_x\left(p_0\frac{c}{a}-\epsilon\frac{\sqrt{p_0^2-a}}{a}\sqrt{ab+c^2}\right) \;.
\ee
Assuming a horizon at $x=0$ and $a(x)=2\kappa x+\mathcal{O}(x^2)$, where $\kappa$ is a constant, the integral picks up an imaginary part from $i\pi$ (it is a semicircle) times the residue at $x=0$,
\be
\mathcal{A}=\frac{\pi p_0}{\epsilon\kappa}[\sqrt{[\epsilon c(0)]^2}-\epsilon c(0)] \;.
\ee
If we assume $\kappa>0$, $c(0)<0$ and $\epsilon=1$, then
\be\label{Akappa}
\mathcal{A}=\frac{2\pi p_0}{\kappa}[-c(0)] \;.
\ee
For~\eqref{ds2GP} with $v(x)$, we find $\mathcal{A}=2\pi p_0/\kappa$, where $\kappa=v'(0)$, which agrees with~\cite{Volovik:1999fc,Unruh:2004zk}. For a Schwarzschild metric, $v=-\sqrt{2M/r}$, we have $v'(r=2M)=1/4M$ and $\mathcal{A}=8\pi M p_0$, which is Hawking's result~\cite{Hawking:1975vcx}. 

\begin{figure*}
    \centering
    \includegraphics[width=0.329\linewidth]{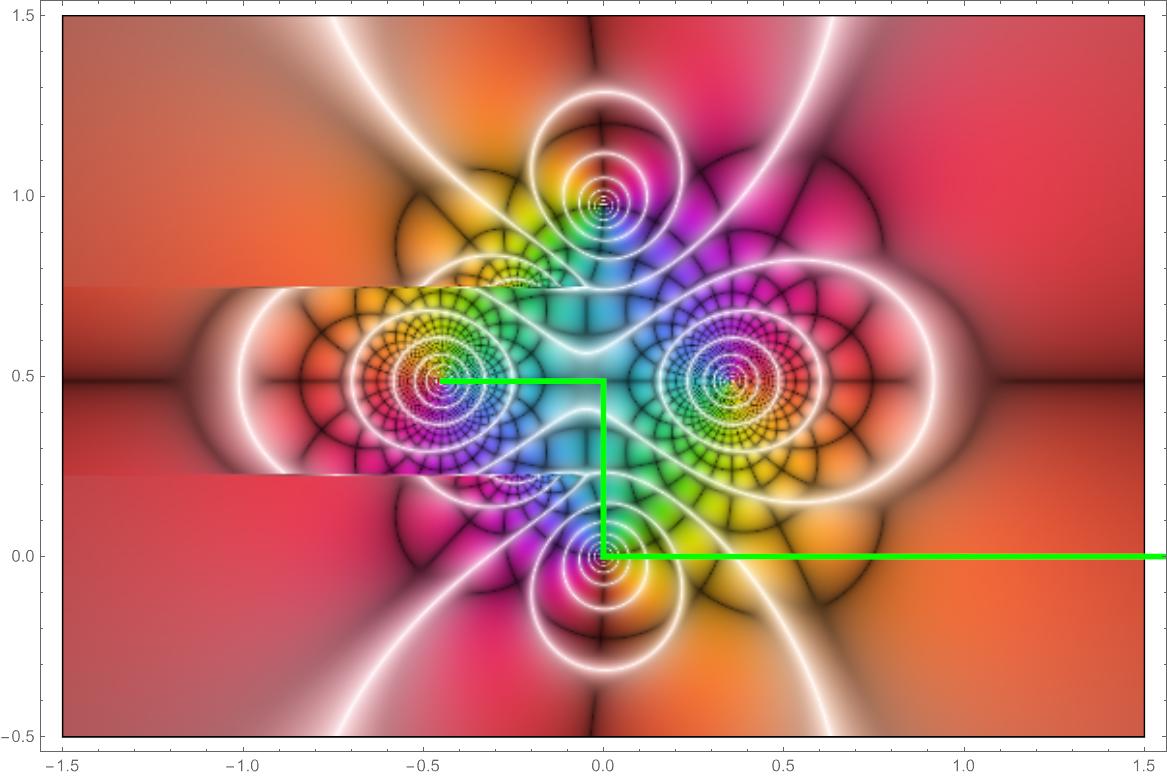}
    \includegraphics[width=0.329\linewidth]{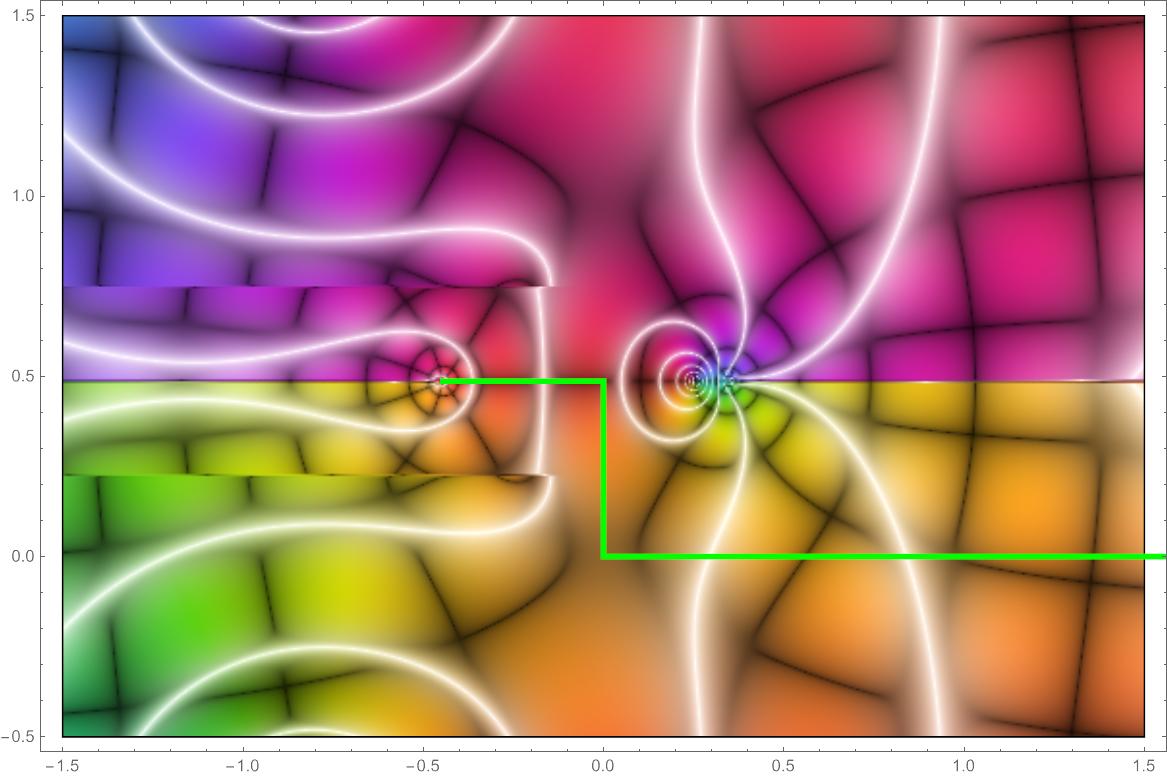}
    \includegraphics[width=0.329\linewidth]{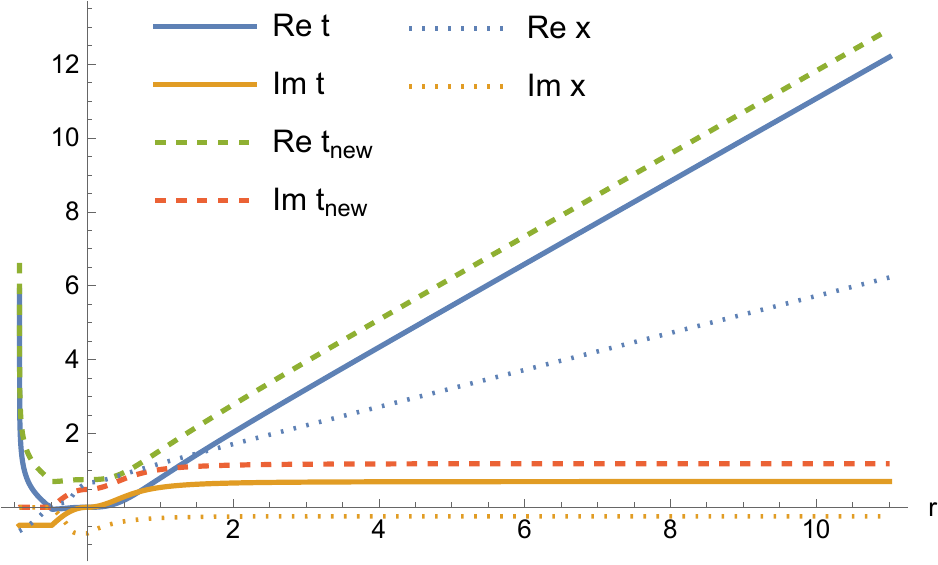}
    \caption{$\dot{t}(u)=\dot{t}_{\rm new}(u)$ and $t_{\rm new}(u)$ for the case in Fig.~\ref{fig:xComplexPlot3start}, but with the origin of $u$ chosen as $\dot{t}(0)=0$. The green contour is equivalent to the one in Fig.~\ref{fig:xComplexPlot3start}. The third plot shows $t$, $t_{\rm new}$, $x$ along this contour.}
    \label{fig:txPGomega0}
\end{figure*}

We could obtain~\eqref{Akappa} for $g_{\mu\nu}(x)$ without having to find the instanton, but this is not possible for $g_{\mu\nu}(t,x)$. Since we will use the results for $g_{\mu\nu}(x)$ as a starting point in a numerical continuation towards $g_{\mu\nu}(t,x)$, we will go back and consider the instanton in~\eqref{dtdxabcStat} in more detail, focusing on GP, where 
\be\label{dtdxGPstat}
\dot{t}=\frac{\sqrt{1+p^2}-v\epsilon\sqrt{p^2+v^2}}{1-v^2}
\qquad
\dot{x}=\epsilon\sqrt{p^2+v^2(x)}
\ee
and
\be\label{AGPstat}
\mathcal{A}=2\text{Im}\int\ud x\frac{\sqrt{1+p^2}v-\epsilon\sqrt{p^2+v^2}}{1-v^2} \;.
\ee
Since $v<0$ for $\text{Im }x=0$, we see that there is only a nonzero residue if $\epsilon=+1$. Eq.~\eqref{dtdxGPstat} agrees with Eq.~(10) in~\cite{Coutant:2012zh} (to translate from their notation, replace $p\to\dot{x}-v\dot{t}$ and $\omega\to\sqrt{1+p^2}$).

We choose to start at~\eqref{dt0dxi}, which gives an equation for $x(0)$,
\be
v[x(0)]=-i\epsilon p_0 \;.
\ee
To observe a particle at $x\to+\infty$, we need $\epsilon=1$.

As an example, we consider
\be\label{vPulse}
v(t,x)=-\frac{\alpha}{(1+[kx]^2)^2}\text{sech}^2(\omega t) \;,
\ee
where for now $\omega=0$. There are two solutions to $v(x=\pm x_H)=-1$ for $\alpha>1$: an outer ($x=x_H>0$) and an inner ($x=-x_H$) horizon. The inner horizon can be thought of as a white-hole horizon for the region $x<-x_H$. By rescaling $q^\mu\to q^\mu/k$ and $u\to u/k$, we find $\mathcal{A}(\omega,k)=H(\omega/k)/k$ for some function $H$. In Sec.~\ref{inflation section} we focused on $p=p_s$ and $p'=p'_s$. For the black-hole types of metrics we have in mind in this section, we consider the case where one particle is observed with momentum $p\ne p_s$, while the other particle is not observed. Performing the integral over $p'$ with the saddle-point method sets $p'=p'_s$. The reason for treating the two particles differently is suggested by the popular description of Hawking radiation as a process where one particle falls into the black hole while the other escapes. A detector outside the black hole, i.e. at $x\to+\infty$, can only detect particles with $\dot{x}(\infty)>0$, but there might not be any saddle point for $p$ there.

\begin{figure*}
    \centering
    \includegraphics[width=0.329\linewidth]{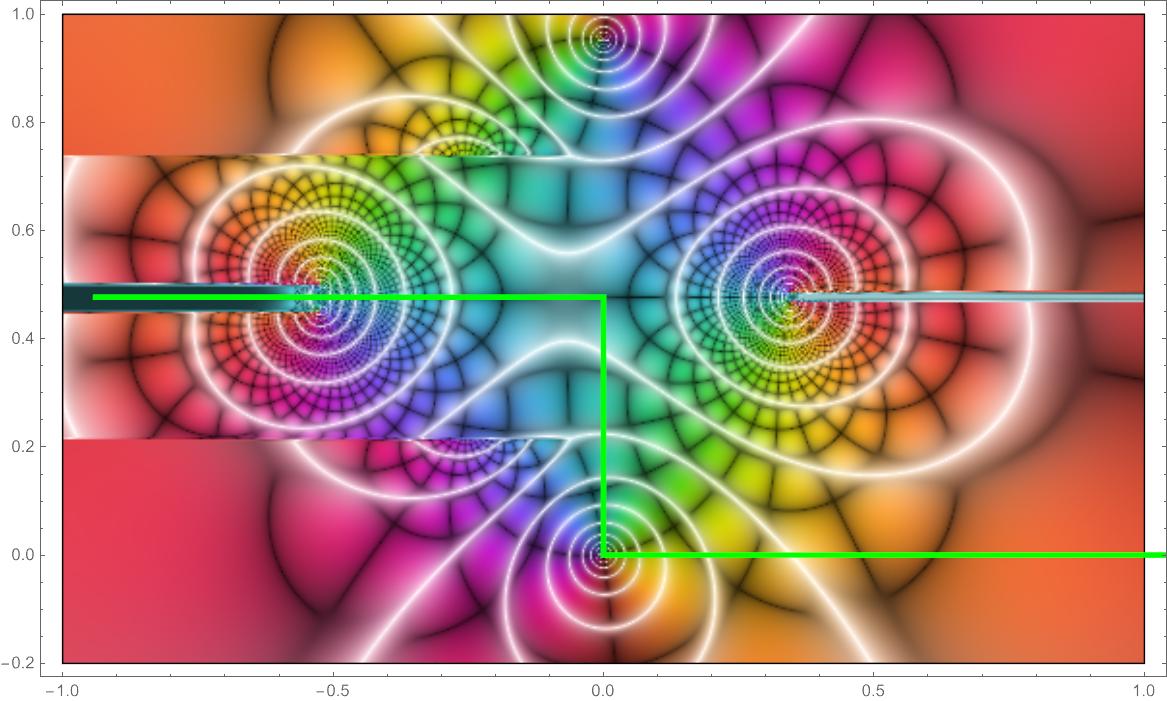}
    \includegraphics[width=0.329\linewidth]{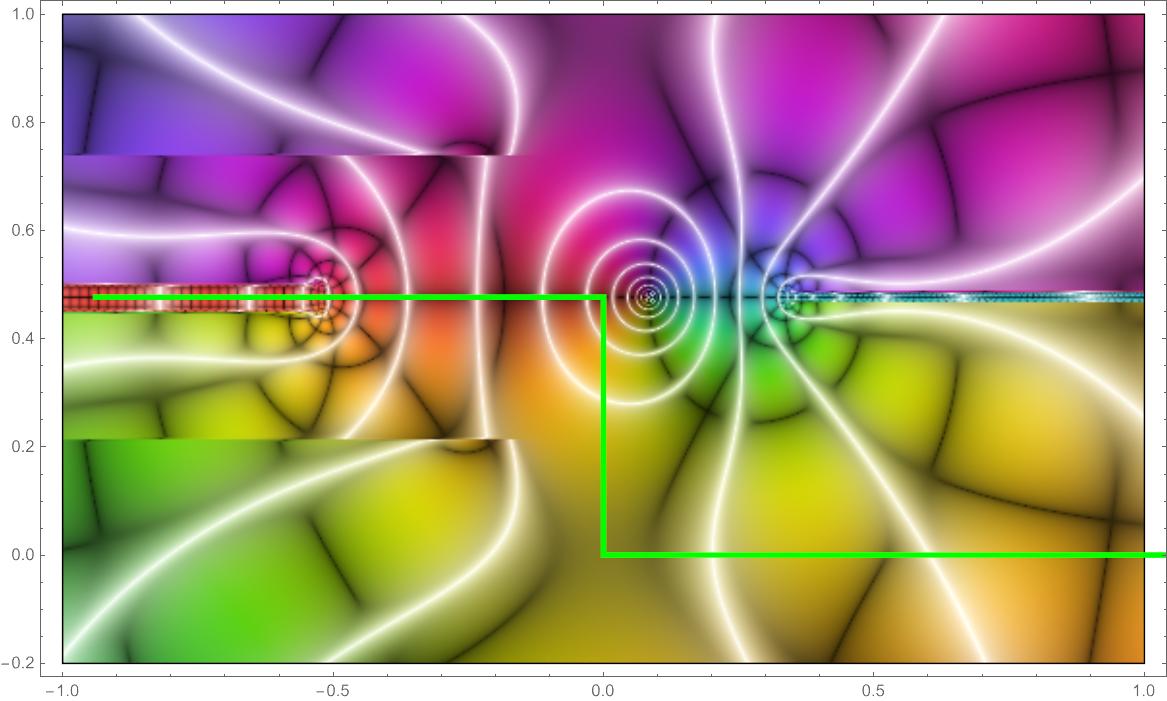}
    \includegraphics[width=0.329\linewidth]{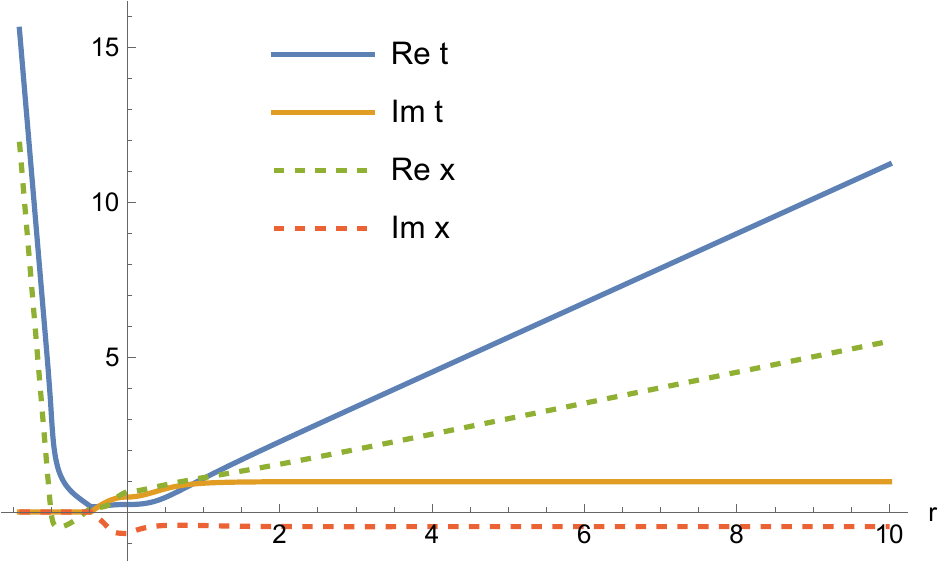}
    \caption{$\dot{t}(u)$ and $t(u)$ for the case in Fig.~\ref{fig:txPGomega0}, but with $\omega=0.4$.}
    \label{fig:txPGomega04}
\end{figure*}

We start arbitrarily at $p^1=0.5$ (so $\epsilon=1$). To see what $u$ contour to choose, we study complex plots as in Sec.~\ref{inflation section}. Since the EOMs do not depend explicitly on $u$, we can without loss of generality start at $x(0)=0$. Fig.~\ref{fig:xComplexPlot3start} shows three potentially relevant contours. From~\eqref{dtdxGPstat} we see that we always have $\dot{t}=\sqrt{1+p^2}>1$ as $|u|\to\infty$ for any $\text{arg }u$. We can also see this in the first plot in Fig.~\ref{fig:xComplexPlot3start}, where there is the same, red color in all asymptotic directions. So for the observed particle to go to the asymptotic future, $t(u)\to+\infty$, that end of the instanton should go to $\text{Re }u\to+\infty$. We would in general expect the other end of the contour to go to $\text{Re }u\to-\infty$, but that is not possible in this case, because then $t\to-\infty$, but there should not be any particles in the asymptotic past. Instead, the contour will stop at $u=u_I\approx-0.4$, where $x(u_I)=x_I\approx-0.64$ is the inner horizon. In other words, the unobserved particle will get stuck at the inner horizon. We can see from either~\eqref{dtdxGPstat} or from Fig.~\ref{fig:xComplexPlot3start} that $|\dot{t}|\to\infty$ at $x_I$. As we will see, for any nonzero $\omega>0$, the instanton will eventually escape from $x_I$.

We have chosen $\epsilon=+1$ to observe a particle with $\dot{x}(\infty)>0$, but as $\text{Re }u\to+\infty$, the asymptotic momentum can be either $\dot{x}(\infty)=+|p|$ or $\dot{x}(\infty)=-|p|$ depending on which direction the $u$ contour goes around the branch points. We can see this in the $\dot{x}$ plots in Fig.~\ref{fig:xComplexPlot3start}, where red means $\dot{x}>0$ and blue $\dot{x}<0$. The solid contour therefore gives a particle with $\dot{x}(\infty)>0$, while the dashed and dot-dashed contours give $\dot{x}(\infty)<0$. We will focus on two of them. Thus, the same values of $q^\mu(0)$ and $\dot{q}^\mu(0)$ can describe different processes depending on how we navigate around the branch points.   

We can compute $\mathcal{A}$ for the three contours conveniently using~\eqref{generalExpFin0} or~\eqref{generalExpFin}, but to compare with previous studies for Hawking radiation by static metrics, we will also compute it using~\eqref{AGPstat}. Fig.~\ref{fig:Agrand3start} shows the integrand in the complex $x$ plane, and $x(u)$ for the three contours in Fig.~\ref{fig:xComplexPlot3start}. The solid $x$ contour, which goes to $x\to+\infty$, can be deformed to the real axis, except for a semi-circle around the horizon at $x=x_H\approx0.64$. The residue theorem then gives the usual Hawking result~\eqref{Akappa}. Such residue calculations can be found in many papers; see e.g.~\cite{Srinivasan:1998ty,Parikh:1999mf,Kim:2007ep}. In contrast, the two contours which go to $x\to-\infty$ give a nonzero imaginary part because they wrap around a branch point. Note that, while it might look like the dashed contour hits the poles at both $x=x_H$ and $x=x_I=-x_H$, the contour passes those points on the Riemann sheet without poles, as seen in the second plot in Fig.~\ref{fig:xComplexPlot3start}. For such contours, we can rewrite~\eqref{AGPstat} as      
\be\label{AGPstatBranch}
\mathcal{A}=4\text{Im}\int_0^{x_B}\ud x\frac{\sqrt{p^2+v^2}}{1-v^2} \;,
\ee
where $x_B$ is the branch point around which the contour goes. \eqref{AGPstatBranch} is a different from~\eqref{Akappa}. Thus, for $\dot{x}(\infty)>0$ we find the usual residue integral (i.e. a local result) and the usual Hawking result, while for $\dot{x}(\infty)<0$ we find a non-local integral around a branch point. The fact that $\dot{x}(\infty)<0$ means that both particles fall into the black hole probably explains why contributions such as \eqref{AGPstatBranch} has not been studied before (as far as we are aware). However, we will show that studying such contributions can help to resolve some questions about previous methods and ideas for studying $\dot{x}(\infty)>0$. Also, in more general analog systems, one might be able to observe even ``in-falling'' particles, e.g. as waves emitted from an analog white-hole horizon~\cite{Rousseaux:2007is,Macher:2009tw,Macher:2009nz,Rousseaux:2010md,Weinfurtner:2010nu,Mayoral:2010ck,Fourdrinoy:2021hmi}.     

\begin{figure*}
    \centering
    \includegraphics[width=0.329\linewidth]{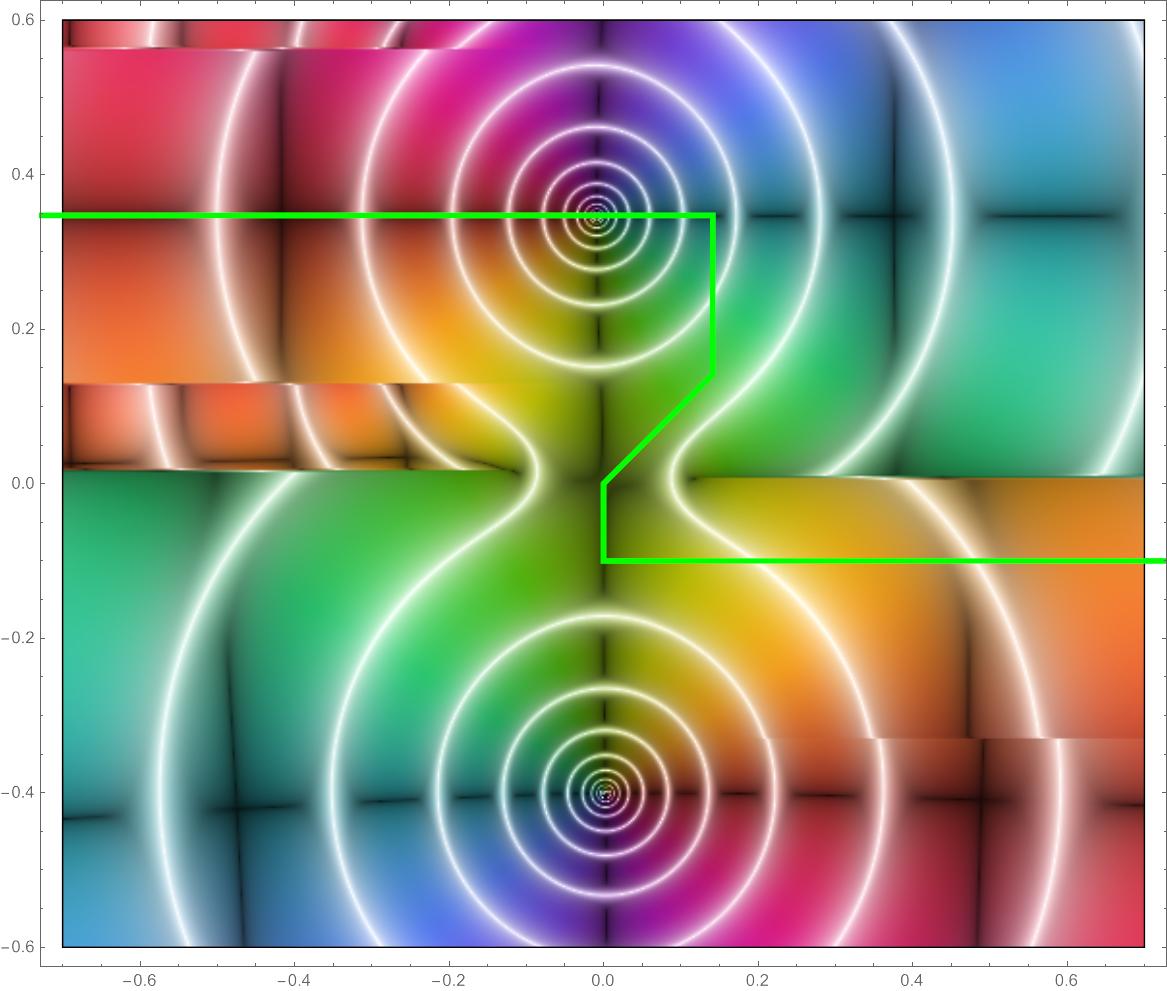}
    \includegraphics[width=0.329\linewidth]{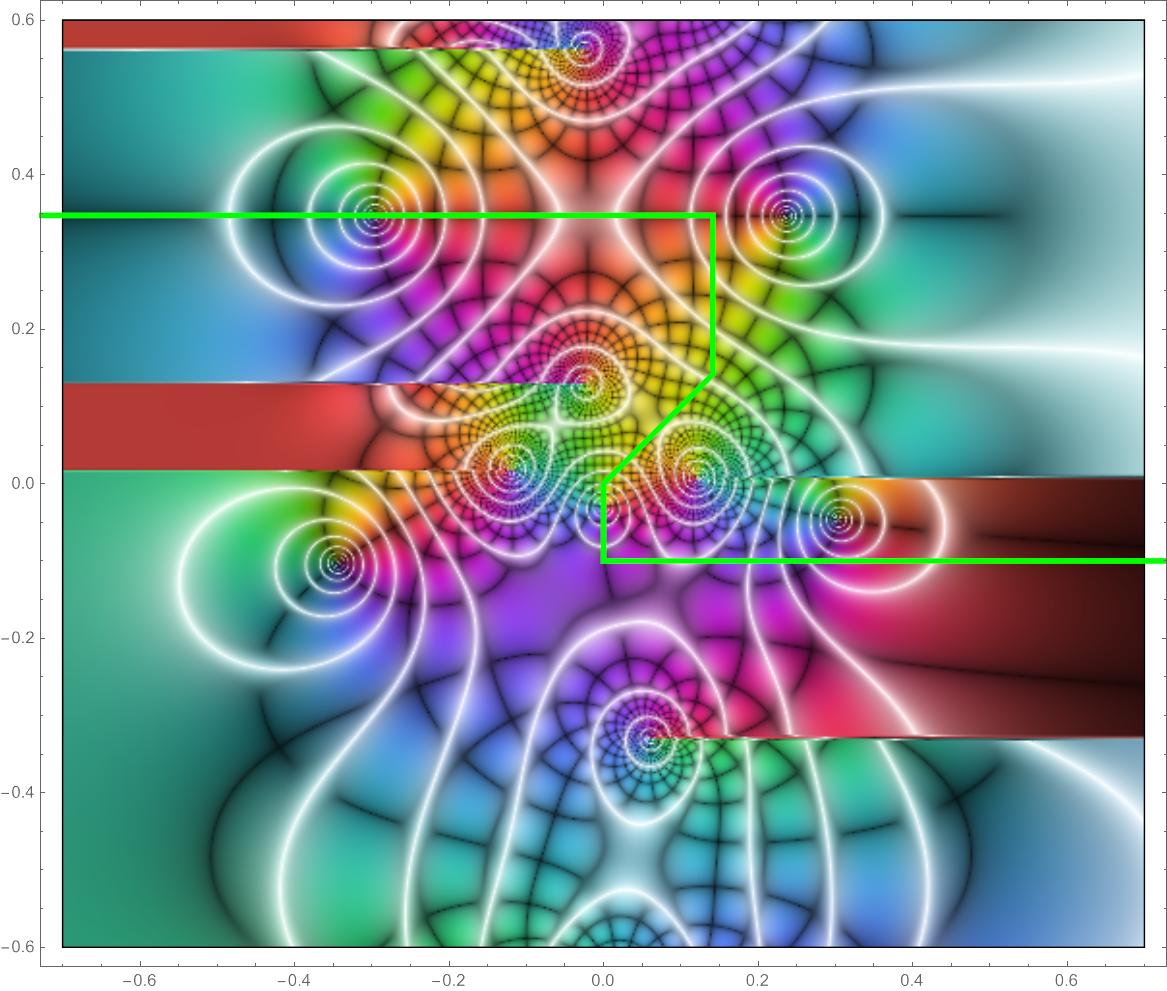}
    \includegraphics[width=0.329\linewidth]{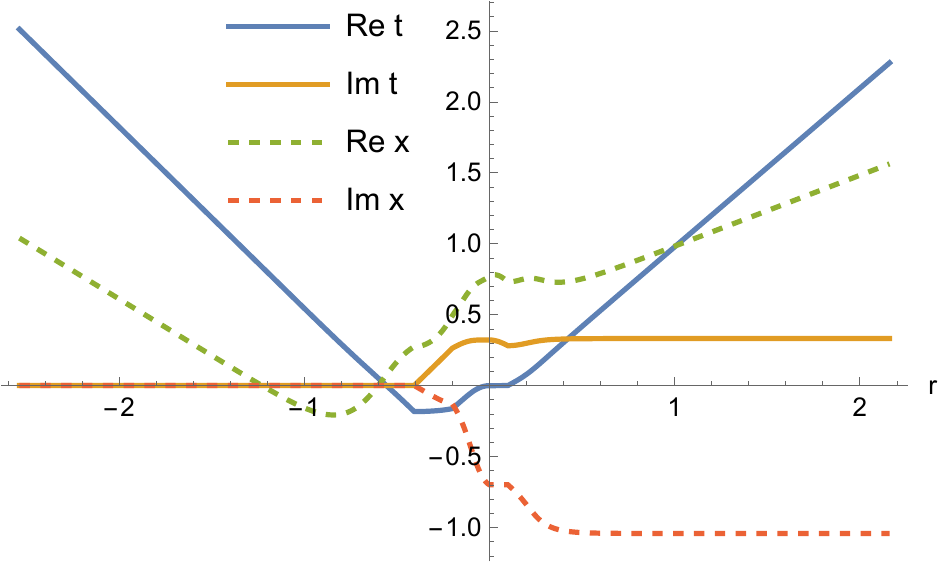}
    \caption{$t(u)$ and $\dot{x}(u)$ for the case in Fig.~\ref{fig:txPGomega04}, but with $\omega=3.35$.}
    \label{fig:txPGomega3p35}
\end{figure*}

\subsection{The first step in a numerical continuation}

Since $g_{\mu\nu}(x)$ does not depend on $t$, it might at first seem like the static limit considered in the previous section cannot suggest a starting point for $t(0)$ for $g_{\mu\nu}(t,x)$, but it is actually possible using ideas from~\cite{DegliEsposti:2023qqu,DegliEsposti:2024rjw}. Consider a metric $v(t,x)=v(x)w(\omega t)$, where\footnote{The coefficient of $\mathcal{O}(\omega^2)$ can be seen as a definition of $\omega$.} $w=1-(\omega t)^2+\mathcal{O}(\omega^3)$. The $\mathcal{O}(\omega^2)$ correction to~\eqref{Akappa} can be obtained without finding the correction to the instanton, by using~\eqref{udToPartial} with $p_x\to\omega^2$. We find $\mathcal{A}\approx\eqref{Akappa}+\delta\mathcal{A}$, where
\be\label{deltaAt2}
\delta\mathcal{A}=-2\omega^2\text{Im}\int\ud u\, t^2 v(x)\dot{t}[v(x)\dot{t}-\dot{x}] \;,
\ee
where $q^\mu$ is the instanton for $\omega=0$. The fact that~\eqref{deltaAt2} depends on an undetermined constant, $t(0)$, might at first seem like a problem, but demanding that the value of $t(0)$ should maximize~\eqref{deltaAt2} gives a condition that allows us to determine $t(0)$.  
However, \eqref{deltaAt2} is only one real equation, so we need another condition to determine both $\text{Re }t(0)$ and $\text{Im }t(0)$. 
Fig.~\ref{fig:txPGomega0} shows a $u$ contour which is equivalent to the solid contour in Fig.~\ref{fig:xComplexPlot3start}, but with the segment going into the inner horizon included (and with the origin chosen such that $\dot{t}(0)=0$). From the third plot in Fig.~\ref{fig:txPGomega0} we see that $\text{Im }t(u)=:-V$ is constant but nonzero along this segment, while $\text{Im }x(u)=\text{Im }\dot{x}(u)=\text{Im }\dot{t}(u)=0$. Since the value of $t(0)$ is irrelevant for $\omega=0$, we can define another solution $t_{\rm new}(u)=t(u)+U+iV$, where $\text{Im }U=0$, so that $\text{Im }t_{\rm new}(u)=0$ on that segment. $U$ is determined by plugging $t\to t_{\rm new}$ in~\eqref{deltaAt2} and setting $\partial_U\delta\mathcal{A}=0$,
\be\label{Usol}
U=-\frac{b}{a}
\qquad
\{a,b\}=\text{Im}\int\ud u\{1,t+iV\}v\dot{t}[v\dot{t}-\dot{x}] \;.
\ee
Thus, while $t(0)$ is irrelevant for $\omega=0$, for $\omega>0$ we have $\lim_{\omega\to0}t(0)=U+iV\ne0$, so $t(0)=U+iV$ is a better starting point than $t(0)=0$ for a numerical continuation starting at $\omega=0$. We avoided a similar problem in Sec.~\ref{inflation section} by considering symmetric metrics and $p=p_s$, $p'=p'_s$, for which $x(0)=0$ for any value of $\beta$. For the particular example we consider in Fig.~\ref{fig:txPGomega0} we have $t_{\rm new}(0)\approx0.75+0.48i$.  

\begin{figure}
    \centering
    \includegraphics[width=\linewidth]{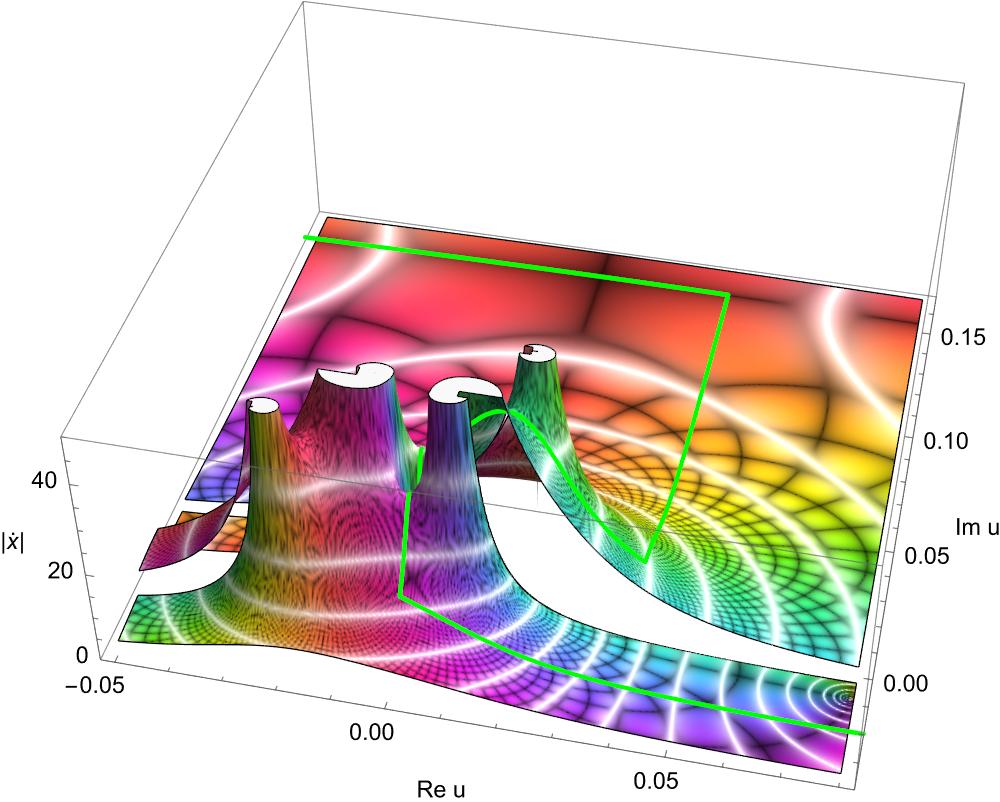}
    \caption{$\dot{x}(u)$ for the case in Fig.~\ref{fig:txPGomega04}, but with $\omega=10$. The truncated peaks are poles.}
    \label{fig:dxComplexPlot3DGPomega10}
\end{figure}

\begin{figure}
    \centering
    \includegraphics[width=\linewidth]{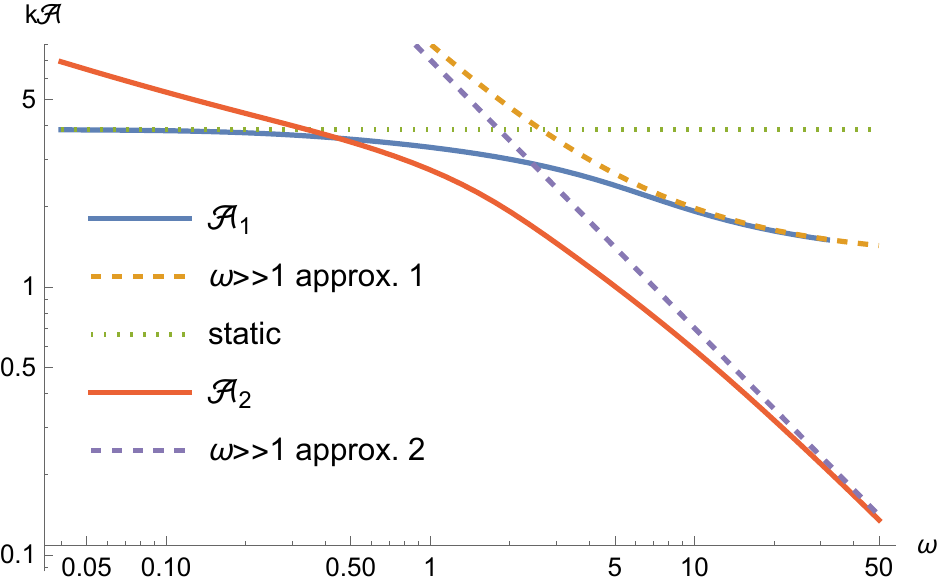}
    \caption{$\mathcal{A}(\omega)$ for~\eqref{vPulse} with $k=1$, $\alpha=2$, $p^1=0.5$ and $p'=p'_s$ (or $p^{\prime1}=0.5$ and $p=p_s$). Instantons for $\mathcal{A}_1$ for a couple of different values of $\omega$ are shown in Figs.~\ref{fig:txPGomega0}, \ref{fig:txPGomega04}, \ref{fig:txPGomega3p35} and~\ref{fig:dxComplexPlot3DGPomega10}. The ``static'' line is Hawking's result~\eqref{Akappa}. The first $\omega\gg1$ approximation is given by~\eqref{AlargeOmega}, while the second is given by~\eqref{AperturbativeLimit}.}
    \label{fig:AGPomega0to30}
\end{figure}

\subsection{Numerical continuation to $\omega>0$}

Using the results from the previous subsection, we now start a numerical continuation from $\omega=0$ to $\omega>0$. We have chosen a step size of $\Delta\omega=0.05$. The instanton for $\omega=0.4$ is shown in Fig.~\ref{fig:txPGomega04}. By comparing Figs.~\ref{fig:txPGomega0} and~\ref{fig:txPGomega04}, we see that the branch and singular point at the inner (and outer) horizon for $\omega=0$ splits into two branch points, and a region opens up between them where the instanton can escape the black hole. In other words, we can now integrate along the upper horizontal line to $\text{Re }u\to-\infty$ without encountering any branch points or poles, and $t\to+\infty$ in this region. As we increase $\omega$ further, the branch points move around and from time to time we need to adjust the $u$ contour in order to stay on the same side of the branch points and poles. One such example is illustrated in Fig.~\ref{fig:txPGomega3p35}. The two plots over the complex $u$ plane illustrate a typical situation, where it is easier to see how to choose a $u$ contour by looking at $\dot{x}(u)$ rather than $t(u)$. As $\omega$ becomes large, it becomes increasingly difficult to compute the instanton, because it is squeezed between converging poles, as shown in Fig.~\ref{fig:dxComplexPlot3DGPomega10}.     
    
The results for $\mathcal{A}(\omega)$ is shown in Fig.~\ref{fig:AGPomega0to30}. Since we used $\omega=0$ as a starting point, it is not surprising that $\lim_{\omega\to0}\mathcal{A}(\omega)=\eqref{Akappa}$ agrees with Hawking's result. It is not obvious, though, what to expect in the opposite limit, $\omega\gg1$.  

\begin{figure}
    \centering
    \includegraphics[width=\linewidth]{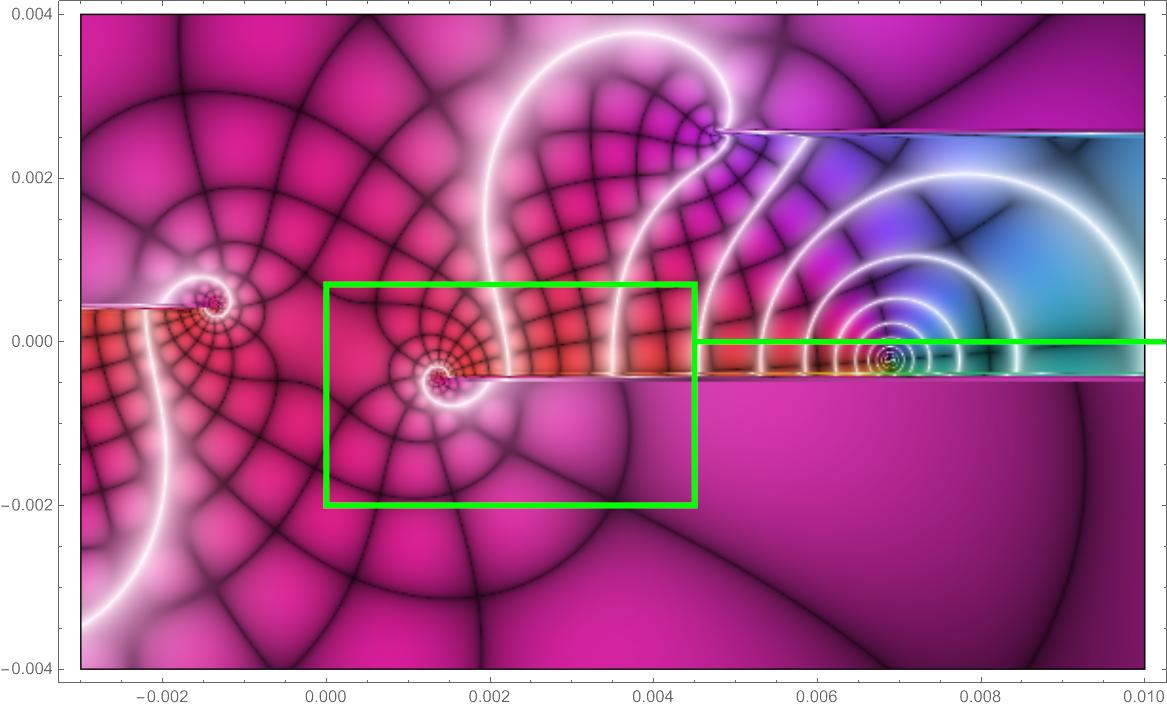}
    \includegraphics[width=\linewidth]{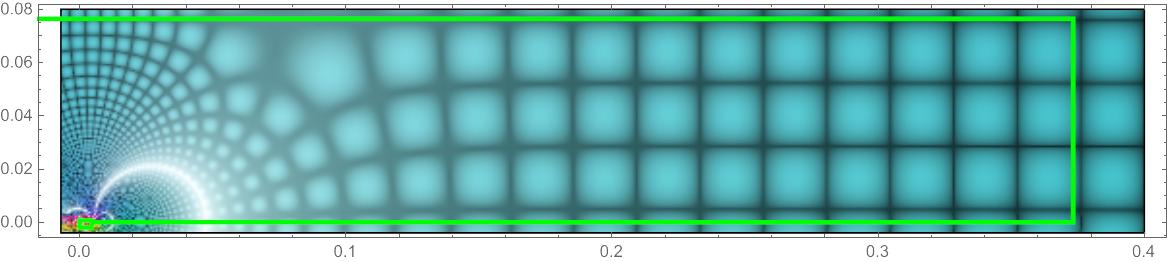}
    \includegraphics[width=.9\linewidth]{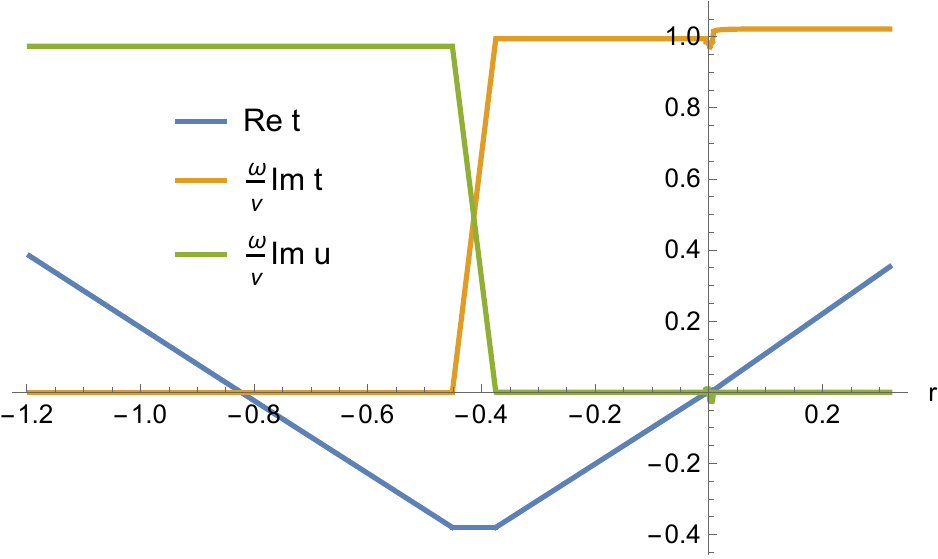}
    \includegraphics[width=.9\linewidth]{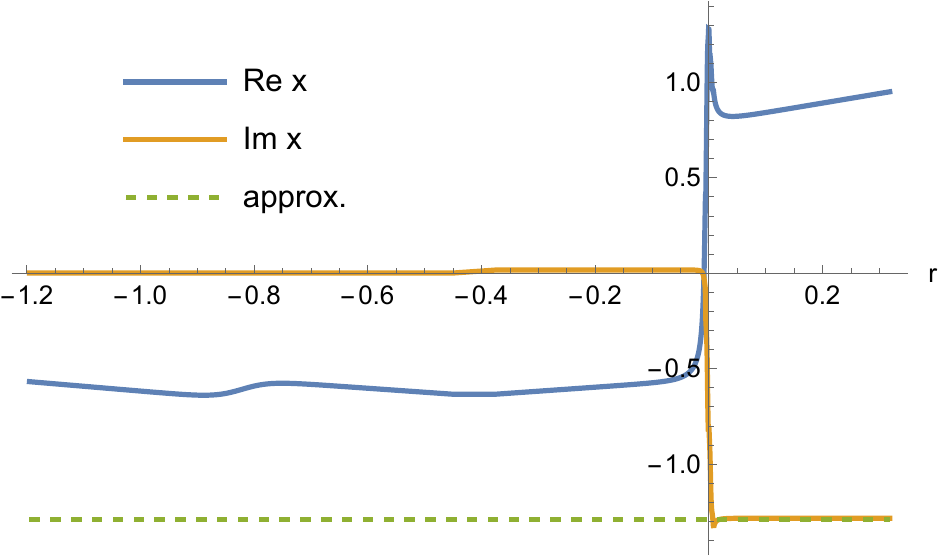}
    \includegraphics[width=.9\linewidth]{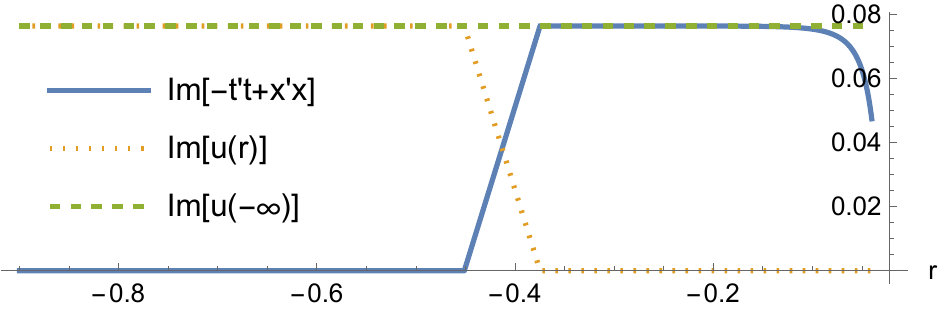}
    \caption{$x(u)$ for the case in Fig.~\ref{fig:txPGomega04}, but with $\omega=20$. $\nu=\pi/2$ for a Sauter pulse. (The two ends of the $u$ contour go to $\text{Re }u\to-\infty$ and $\text{Re }u\to+\infty$, but the code stops automatically when $\ddot{t}$, $\ddot{x}$ or $v$ is sufficiently small, and in this example it stops somewhere $\text{Re }u<3.9$, so one cannot see that end of the contour in the first plot.)}
    \label{fig:LargeOmegaUp}
\end{figure}

\subsection{$\omega\gg1$ approximation}

The instantons we obtained in the previous subsection show that $\dot{x}$ becomes large near $u=0$ as $\omega$ increases, while $\dot{t}$ stays $\mathcal{O}(1)$. To LO we can therefore neglect the square root in~\eqref{dxFromSqrt}. For metrics of the form $v(t,x)=v(x)w(\omega t)$, we then have 
\be\label{eqLeadingToJ}
\frac{\dot{x}}{v(x)}\approx w(\omega t)\dot{t} \;.
\ee
Writing $w(y)=f'(y)$ ($f(y)=\tanh(y)$ for a Sauter pulse) and defining
\be
J(x):=\int_0^x\frac{\ud\tilde{x}}{v(\tilde{x})} \;,
\ee
we have
\be\label{JminusJ}
J[x(u)]-J[x(u')]\approx\frac{1}{\omega}(f[\omega t(u)]-f[\omega t(u')]) \;,
\ee
where $u$ and $u'$ are two different points on the contour. Combining the $\ddot{t}$ equation in~\eqref{GPeqs} and~\eqref{eqLeadingToJ} gives a second equation,
\be
\frac{\dot{t}\ddot{t}}{1-\dot{t}^2}\approx\frac{v'(x)}{v(x)}\dot{x} \;,
\ee
which integrates to
\be\label{v2overv2}
\frac{v^2[x(u)]}{v^2[x(u')]}\approx\frac{1-\dot{t}^2(u')}{1-\dot{t}^2(u)} \;.
\ee

We consider a $u$ contour obtained by pulling the upper vertical segment in Fig.~\ref{fig:dxComplexPlot3DGPomega10} towards $\text{Re }u\to\infty$, so that the lower half wraps around the branch cut in Fig.~\ref{fig:dxComplexPlot3DGPomega10} which goes to $\text{Re }u\to\infty$; see Fig.~\ref{fig:LargeOmegaUp}. From~\eqref{generalExpFin0} and the last plot in Fig.~\ref{fig:LargeOmegaUp}, we see that for this type of contour we have
\be\label{AlargeOmega1}
\mathcal{A}=2\text{Im}[pq(u_1)+u_0]=2\text{Im}[pq(u_1)-\dot{q}(\tilde{u}_0)q(\tilde{u}_0)] \;,
\ee
where, for this particular example, we can choose $\tilde{u}_0=u(r=-0.2)$. We denote $x(\tilde{u})=x_{\rm in}$ and $x(\tilde{u})=x_{\rm out}$ for $r<0$ and $r>0$, and similarly for $t$. We see in Fig.~\ref{fig:LargeOmegaUp} that $\text{Im }t_{\rm in}\approx\text{Im }t_{\rm out}\approx\nu/\omega$ ($\nu=\pi/2$ for this example). Since $p_1'=p_{1s}'$ is the saddle-point momentum, we can obtain $\text{Im }x_{\rm in}$ from $\text{Im }t_{\rm in}$ using~\eqref{saddleCondition},  
\be\label{saddleConditionLargeOmega}
\text{Im }x_{\rm in}=\frac{\dot{x}_{\rm in}}{\dot{t}_{\rm in}}\text{Im }t_{\rm in}=-\frac{\tilde{p}_1}{\sqrt{1+\tilde{p}_1^2}}\text{Im }t_{\rm in} \;.
\ee 
We have four real constants, $\text{Re }x_{\rm in}$, $\tilde{p}^1$, $\text{Re }x_{\rm out}$ and $\text{Im }x_{\rm out}$, which we can determine using~\eqref{JminusJ} and~\eqref{v2overv2},
\be\label{Jandv}
J(x_{\rm out})\approx J(x_{\rm in})
\qquad
\frac{v(x_{\rm out})}{v(x_{\rm in})}\approx\frac{\dot{x}_{\rm in}}{\dot{x}_{\rm out}}=-\frac{\tilde{p}^1}{p^1} \;.
\ee
Thus, we can approximate~\eqref{AlargeOmega1} as
\be\label{AlargeOmega}
\mathcal{A}\approx2\left[\frac{\nu}{\omega}[\sqrt{1+p_1^2}+\sqrt{1+\tilde{p}_1^2}]+p_1\text{Im }x_{\rm out}-\tilde{p}_1\text{Im }x_{\rm in}\right] \;.
\ee
Note that~\eqref{AlargeOmega} does not depend on $\alpha$. $\mathcal{A}$ converges to~\eqref{AlargeOmega} for both $\alpha\ll1$ and $\omega\gg1$.

For $\omega\gg1$ we can simplify~\eqref{AlargeOmega} further by noting that, since $x_{\rm in}$ is a saddle point, we have $\mathcal{A}(x_{\rm in}+\delta)=\mathcal{A}(x_{\rm in})+\mathcal{O}(\delta^2)$, which means that we can neglect $\text{Im }x_{\rm in}$ in~\eqref{Jandv} and~\eqref{AlargeOmega}, since~\eqref{saddleConditionLargeOmega} shows that $\text{Im }x_{\rm in}=\mathcal{O}(\omega^{-1})$. Then $\text{Im }kx_{\rm out}$ and $\tilde{p}_1/p_1$ only depend on the pulse shape, but not on $\alpha$, $k$ or $\omega$. For~\eqref{vPulse} we find $\text{Im }kx_{\rm out}\approx-1.29$ and $\tilde{p}_1/p_1\approx0.522$. In Fig.~\ref{fig:AGPomega0to30} we see that this approximation agrees well with the exact result for $\alpha=\mathcal{O}(1)$ and $\omega\gg1$. However, it is not a trivial generalization of~\eqref{perturbativeFourier}, which one might naively have expected. To understand why, we have to consider the 2D $p-q$ momentum space in more detail.

\begin{figure}
    \centering
    \includegraphics[width=\linewidth]{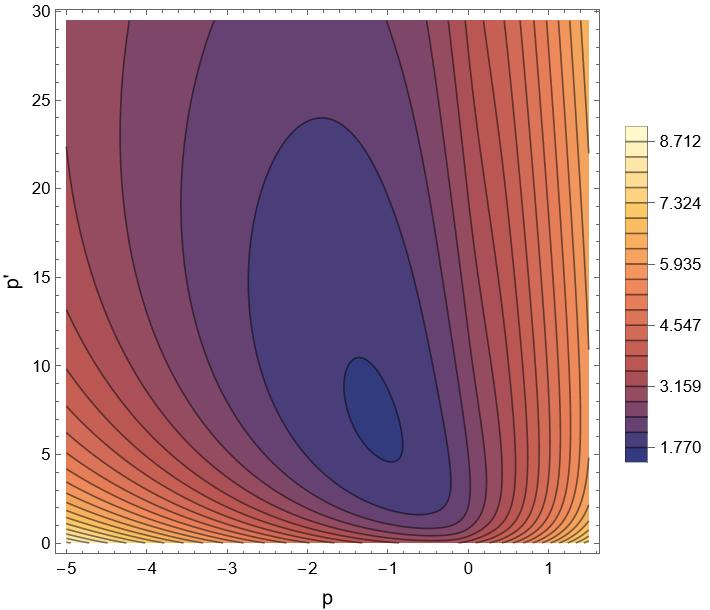}
    \caption{The exponential part of the spectrum, $\mathcal{A}(p=p^1,p'=p^{\prime1})=\mathcal{A}_1(p,p')$, of the two produced particles for the metric in~\eqref{vPulse} for $\alpha=2$, $k=1$ and $\omega=1$. The result for $p<0$ can be obtained with a numerical continuation starting e.g. at $\{p,p'\}\approx\{-1,6.34\}$ with $\{t(0),x(0)\}\approx\{0.0019+0.43i,0.55-0.71i\}$. The result for $p>0$ can be obtained starting at $\{p,p'\}\approx\{1,5.62\}$ with $\{t(0),x(0)\}\approx\{0.019+0.44i,0.57-0.71i\}$. In both cases $\{\dot{t}(0),\dot{x}(0)\}=\{0,i\}$. With $\{\dot{t}(0),\dot{x}(0)\}=\{0,-i\}$ we find $\mathcal{A}(p,p')=\mathcal{A}_2(p,p')=\mathcal{A}_1(p',p)$, as explained after~\eqref{dt0dxi}, so the full spectrum is symmetric in $p\leftrightarrow p'$.}
    \label{fig:pqPlotOmega1}
\end{figure}

\begin{figure}
    \centering
    \includegraphics[width=\linewidth]{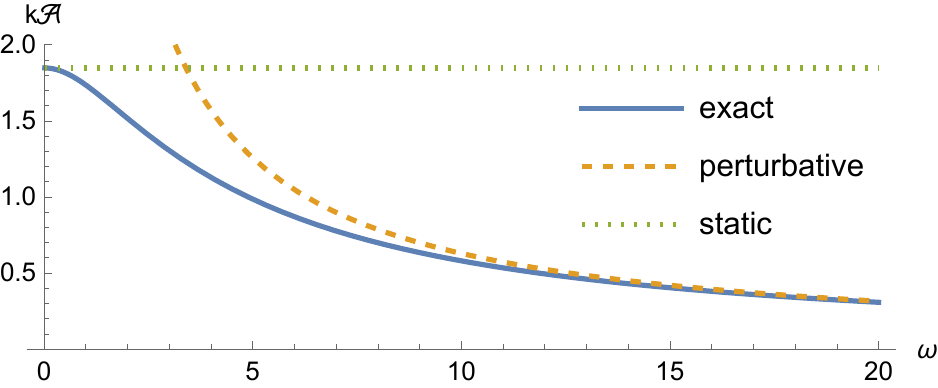}
    \includegraphics[width=\linewidth]{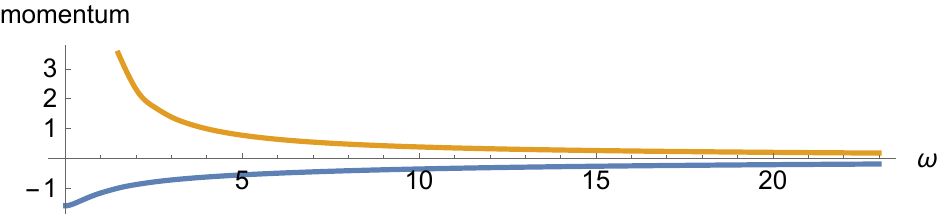}
    \caption{$\mathcal{A}(\omega)$, $p^1=p_s^1(\omega)$ and $p'=p^{\prime1}_s(\omega)$ for the metric in~\eqref{vPulse} with $\alpha=2$, $k=1$. The static result is given, not by Hawking's result~\eqref{Akappa}, but by~\eqref{AGPstatBranch}. The perturbative line is given by~\eqref{AperturbativeLimit}.}
    \label{fig:GPsaddleOmega}
\end{figure}

\subsection{2D momentum space}

Fig.~\ref{fig:pqPlotOmega1} shows $\mathcal{A}(p,p')$ as a function of the momenta of both particles. Obtaining this plot involves some manual work, because we have to choose different $u$ contours in different parts of the momentum space. We compute the instanton for each point on a grid in $(p,p')=(p_m,p'_n)$, where $m$ and $n$ are integers. Then we make an interpolation function with third-order polynomials. We have used Mathematica's Interpolation[...] function. One could interpolate using just the values of $\mathcal{A}(p_m,p'_n)$ at each point, but it is better to incorporate also the gradient, $\nabla\mathcal{A}=(\partial_p\mathcal{A},\partial_{p'}\mathcal{A})$, which we obtain for free by simply evaluating~\eqref{dAdp} and~\eqref{dAdpp}. This approach allows us to use a relatively sparse grid. We have used $|p_{m+1}-p_m|=|p'_{m+1}-p'_m|=0.5$ for Fig.~\ref{fig:pqPlotOmega1}. While we have found $\mathcal{A}(p,p')$ to be continuous at $p=0$ or $p'=0$, $\nabla\mathcal{A}$ is not, so we make separate interpolation functions for $p<0$ and $p>0$. 

The result for $\mathcal{A}_1$ in Fig.~\ref{fig:AGPomega0to30} for $\omega=1$ corresponds to the point $p=0.5$ and $p'=p'_s\approx4.1$ in Fig.~\ref{fig:pqPlotOmega1}. We found $\mathcal{A}_1$ by starting a numerical continuation with the static, Hawking limit, $\omega=0$. In the previous subsection, we found an $\omega\gg1$ approximation for $\mathcal{A}_1$ given by~\eqref{AlargeOmega}, which is independent of $\alpha$ and yet different from the simple perturbative result~\eqref{AperturbativeLimit}. However, we should also consider the point $p'=0.5$ and $p=p_s\approx-0.48$ in Fig.~\ref{fig:pqPlotOmega1}, which corresponds to $\mathcal{A}_2$ for $\omega=1$ in Fig.~\ref{fig:AGPomega0to30}. For $\omega\ll1$, $\mathcal{A}_1$ gives the dominant contribution, which is not surprising since $\lim_{\omega\to0}\mathcal{A}_1=\mathcal{A}_{\rm Hawking}$ and we do not expect to find some contribution which is larger than the Hawking result for $\omega=0$. But at some $\omega=\mathcal{O}(1)$, $\mathcal{A}_2$ becomes smaller than $\mathcal{A}_1$, so for $\omega\gg1$ $\mathcal{A}_2$ gives the dominant contribution, and Fig.~\ref{fig:AGPomega0to30} shows that $\mathcal{A}_2$ converges to the perturbative result~\eqref{AperturbativeLimit} with $p'_s\to-p=-0.5$. In other words, we find convergence to the Hawking result for $\omega\ll1$ and to the perturbative result for $\omega\gg1$, but the two limits are dominated by two different contributions, in contrast to the ``inflation'' example in Fig.~\ref{fig:AofAlpha}. However, in Fig.~\ref{fig:AofAlpha} we let $p=p_s(\alpha)$ and $p'=p'_s(\alpha)$, while in Fig.~\ref{fig:AGPomega0to30} we kept one momentum fixed. 

Consider therefore $p=p_s(\omega)$ and $p'=p'_s(\omega)$, starting the numerical continuation with the saddle point for $\omega=1$, given by the peak in Fig.~\ref{fig:AGPomega0to30}. The results are shown in Fig.~\ref{fig:GPsaddleOmega}. We find $\lim_{\omega\gg1}\mathcal{A}=\mathcal{A}_{\rm perturbative}=\eqref{AperturbativeLimit}$, and the same curve converges to a constant as $\omega\to0$, $\lim_{\omega\to0}k\mathcal{A}=\text{const.}\approx1.85$. However, this constant is not the Hawking result~\eqref{Akappa}, $k\mathcal{A}_H\approx3.45\sqrt{1+p^2}$. It is instead equal to the non-local result in~\eqref{AGPstatBranch} for $\omega=0$, which we mentioned was the result for a process where both particles fall into the black hole. However, in the plot in Fig.~\ref{fig:GPsaddleOmega} of $p_s(\omega)$ and $p'_s(\omega)$, we see that one particle has positive momentum, while the other has negative momentum, for any $\omega>0$. At first it might seem promising that the probability is maximized at such a point, because it is what one might have wanted to see based on the popular description of Hawking radiation as a process where one particle falls into, while the other is emitted from a black hole. However, the agreement with~\eqref{AGPstatBranch} suggests that both particles should fall into the black hole, and we also see that the positive momentum diverges as $\omega\to0$, which is not what one would expect for Hawking radiation. The reason is that the particle with positive momentum is actually the particle that should be trapped by the inner horizon for $\omega=0$. We will now show that this divergence is an artifact.

\subsection{Particles trapped at the inner horizon}

To study the dynamics near the inner horizon, it is convenient to parametrize the instanton using coordinate time, $t$, rather than proper time, $u$, i.e. $\{t(u),x(u)\}\to\{u(t),x(t)\}$. The geodesic equations~\eqref{GPeqs} can be expressed as two separate equations for $x(t)$,
\be\label{xofteq}
x''(t)=([x'-v]^3+v)\partial_x v+\partial_t v \;,
\ee
and for $d(t)=u'(t)$,
\be\label{dofteq}
d'(t)=d(1-d^2)\partial_x v \;.
\ee
A particle that follows the stream, i.e. 
\be\label{streamParticle}
\tilde{x}'(t)=v[t,\tilde{x}(t)] \;,
\ee
is a solution to~\eqref{xofteq}, for which $\tilde{x}'(t)\ne0$ along any real trajectory, for any $\omega$. So $\tilde{x}(t)$ does not describe any trapped particle. However, $\tilde{x}(t)$ is not the solution we encountered e.g. in Fig.~\ref{fig:GPsaddleOmega}. To see the trapped solutions, we consider first the limit $\omega=0$ and expand~\eqref{xofteq} around the inner horizon, $x(t)=x_I+\delta x(t)$, where $v(x_I)=-1$ and $v'(x_I)=-\kappa<0$ (for the outer horizon we have $v'(x_H)=\kappa>0$). To $\mathcal{O}(\delta x)$, we find   
\be
\delta x''+3\kappa\delta x'+2\kappa^2\delta x=0 \;,
\ee
so
\be\label{deltaxSol}
\delta x=a_1 e^{-\kappa t}+a_2 e^{-2\kappa t} \;,
\ee
where $a_1$ and $a_2$ are constants. Thus, $\delta x\to0$ as $t\to\infty$, i.e. the particle is trapped\footnote{Geodesics trapped at a horizon has also been studied in~\cite{Gaur:2023ved}.}.

However, the corresponding solution to~\eqref{dofteq} is
\be\label{dottoftTrapped}
\dot{t}=\frac{1}{d(t)}=\pm\sqrt{1+e^{2\kappa(t-t_0)}} \;,
\ee
where $t_0$ is a constant, so $\dot{t}\to\infty$ as $t\to\infty$, which is why we use $t$ here instead of $u$ to parametrize $x$. Note, though, that if one wants to use the $t$ parametrization not just for this trapped-particle analysis but for a full computation of $\mathcal{A}$, then one in general needs to find $d$ as well as $x$, because $d$ is needed to obtain $\mathcal{A}$.

Eq.~\eqref{deltaxSol} predicts that for $\omega=0$ we have $\lim_{t\to\infty}\delta x=0$, but for any small but nonzero $0<\omega\ll1$ the horizon will eventually disappear, for $\omega t\gg1$, so the trapped particle will eventually escape. One might naively guess that $\omega\ll1$ would mean that we slowly let the particle escape, but the following explains why that is not the case. To see what eventually happens with the trajectory~\eqref{deltaxSol}, we make an ansatz 
\be\label{ansatz}
x'(t)=p+v[t,x(t)] \;,
\ee
which is suggested by~\eqref{streamParticle}. Plugging~\eqref{ansatz} into~\eqref{xofteq} gives 
\be
p(1-p^2)\partial_x v=0 \;,
\ee
so~\eqref{ansatz} is a solution for $p=0$ (which gives~\eqref{streamParticle}) and $p=\pm1$. For $\omega>0$ the particle will approach flat space asymptotically, where the instanton should be a physical trajectory, which means $|x'(t)|<1$ (the speed of light $c=1$), so $p=-1$ is not physical for $v(t,x)<0$. We are left with one candidate,
\be\label{ansatzSol}
x'(t)=1+v[t,x(t)] \;,
\ee
Fortunately, the point $\{x(t_0),x'(t_0)\}=\{x_I,0\}$ is consistent with~\eqref{ansatzSol}. Close to $x_I$ and for $\omega t\ll1$, \eqref{ansatzSol} becomes $x'(t)\approx-\kappa x(t)$, where $\kappa=-\partial_xv(0,x_I)$, which gives $x\approx x_I+a_1e^{-\kappa t}$, which is consistent with~\eqref{deltaxSol} for $t\gg1$. For a while, $x$ will move adiabatically so that $1+v[t,x(t)]\approx0$, but at some $t=t_d$, when $-1<v(t_d,x)<0$ for all $x\in\mathbb{R}$, the horizons disappear, and eventually $v\to0$, so $x'(t)\to1$.

As an aside, \eqref{ansatzSol} is an exact solution for a massless particle, which one can see by reinstating $m$ into~\eqref{dxFromSqrt} and taking $m\to0$.

Thus, consider the $\text{Re u}\to-\infty$ part of the instanton, which is similar to the upper horizontal segment of the $u$ contour in Fig.~\ref{fig:txPGomega04} but for even smaller $\omega$, so that the two branch points are even closer to the contour. As the instanton approaches this pair of branch points, we first have a region where~\eqref{deltaxSol} predicts that the instanton is temporarily trapped near the inner horizon, i.e. $x\approx x_I$. As a function of $t$, $x$ would stay there for a while. But~\eqref{dottoftTrapped} (with $-\sqrt{...}$) shows that $-\dot{t}\gg1$, so $\omega t$ quickly becomes $\mathcal{O}(1)$ when parametrized by $u$, and then one cannot neglect the time dependence, even if $\omega$ is very small. We can instead approximate the instanton with~\eqref{ansatzSol}, which implies $\lim_{t\to\infty}x'(t)\approx1$. Thus, the particle that would have been trapped for $\omega=0$, is thrown out at close to the speed of light for $0<\omega\ll1$. This explains the diverging momentum in Fig.~\ref{fig:GPsaddleOmega}.  

We can see traces of this effect even for $\omega=1$ in Fig.~\ref{fig:pqPlotOmega1}, where the spectrum is quite flat in $p'$ direction. What conclusions should one draw from this? At first, especially if one were only presented with the results in Fig.~\ref{fig:pqPlotOmega1} for $\omega=1$, one might find the flat spectrum promising, because it means a larger probability to produce high-energy particles. But after a closer inspection one would be suspicious, because, looking at the instantons for $\omega=1$, the particle with high energy seems to come from the inner horizon, which might not be what one wanted or expected to see. 

Here we should give a general warning, though. 1) One might be led to think otherwise by the fact that the Hawking's result~\eqref{Akappa} comes from the residue of a pole, but particle production is in general a non-local process, i.e. it has a nonzero formation length.
2) It might be tempting to interpret the points where the instanton $q^\mu(u)$ becomes real as the points where the particles are ``actually'' produced, but for $p\ne p_s$ and $p'\ne p'_s$ all components of $q^\mu(u)$ cannot be real asymptotically simultaneously, and even for $p=p_s$ and $p'=p'_s$ one can choose different $u$ contours so that $q^\mu(u)$ becomes real at different points, or even so that $q^\mu(u)$ never becomes real. Since all continuously deformable contours give exactly the same probability, there is no such thing as the ``correct'' contour. 3) If one wants to study what an observer inside the gravitational field sees, then one would have to think about detectors, which is beyond the scope of this paper.

However, in this case, we find that the contour is squeezed between two branch points which converge as $\omega\to0$, so any equivalent contour would have to go through approximately the same point, where the instanton is close to the inner horizon. So, the suspicion one could have felt from the $\omega=1$ results are confirmed by checking the $\omega\to0$ limit. Thus, the unexpectedly wide spectrum in one direction seems to be due to the inner horizon, or rather the disappearance of it.

(The fact that the initially in-falling particle changes direction reminds us of the black-hole-laser effect found in~\cite{Corley:1998rk}.) 

\begin{figure}
    \centering
    \includegraphics[width=.9\linewidth]{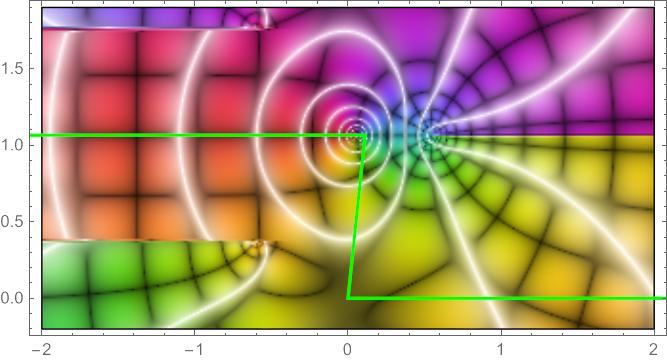}\\
    \includegraphics[width=\linewidth]{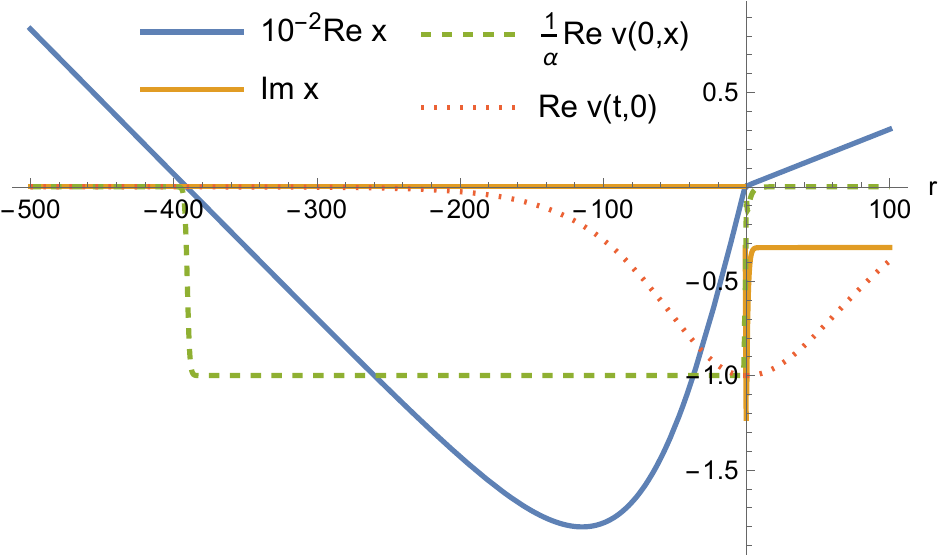}\\
    \includegraphics[width=\linewidth]{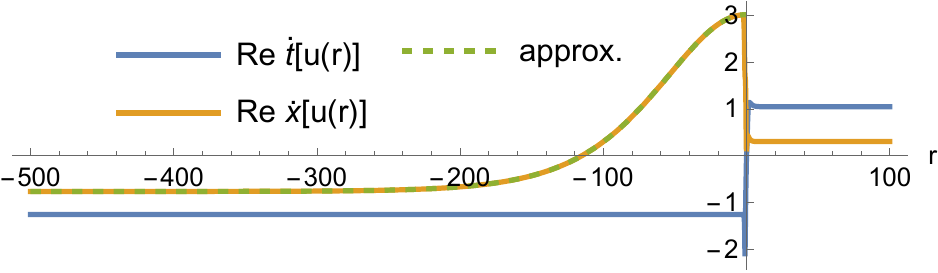}
    \caption{Instanton for~\eqref{vPlateau} with $\alpha=3$, $\omega=0.01$ and $p=0.3$. The approximation is given by~\eqref{plateauApprox1}.}
    \label{fig:plateau}
\end{figure}

\subsection{Avoiding the inner horizon}

If this acceleration by the inner horizon is not what one wants or what one expects for the (analog) black hole one is interested in, then one can avoid it by replacing the spatial pulse shape in~\eqref{vPulse} with one that has a very flat top, e.g. $\tanh(kx)-\tanh(kx+L)$, where $L\gg1$. There are still two horizons, but now we can choose $L$ to be sufficiently large so that the horizons disappear before the instanton has reached the inner horizon. But then we might just as well replace $\tanh(kx)-\tanh(kx+L)\to\tanh(kx)-1$, or some other smooth step. As an example, we consider
\be\label{vPlateau}
v(t,x)=-\alpha\left[1+(\alpha-1)\exp\left(\frac{\alpha}{\alpha-1}x\right)\right]^{-1}\text{sech}^2(\omega t) \;,
\ee
so that we for simplicity have a single horizon at $x=0$ for $\omega=0$, i.e. $v(0,0)=-1$, and $\kappa=\partial_x v(0,0)=1$. Fig.~\ref{fig:plateau} shows the instanton for $\alpha=3$ and $\omega=0.01$. Here $\mathcal{A}$ is well approximated by Hawking's result, $\mathcal{A}\approx\eqref{Akappa}=2\pi\sqrt{1+p^2}$. The second plot shows that one particle is immediately emitted to $x>0$, while the other initially falls into the black hole. The instanton becomes real at $r\sim-1$. There is a long period, until $r\sim-200$, when the instanton experiences an approximately $x$-independent metric\footnote{If $\partial_xv=0$ for all $x$, then there is no pair production.}, $v(t,x)\approx v(t,-\infty)=v(t)$. From~\eqref{xofteq} follows $x'(t)=v(t)+\text{const.}$ and hence
\be\label{plateauApprox1}
\dot{x}(u)=\dot{x}(\tilde{u}_0)+\dot{t}(u)(v[t(u),-\infty]-v[t(\tilde{u}_0),-\infty]) \;,
\ee
where $\tilde{u}_0$ can be chosen as $\tilde{u}_0\sim u(r\lesssim-3)$, so that the in-falling particle has just entered the region where $v(0,x)$ is a plateau. Eq.~\eqref{dofteq} implies that $\dot{t}(u)\approx\dot{t}(-\infty)$ in this region, which we also see in the third plot in Fig.~\ref{fig:plateau}, where $\dot{t}(u)$ is constant even though $\dot{x}(u)$ is not.
 
For $\omega\ll1$ we can simplify~\eqref{plateauApprox1} further by noting that the creation and $\text{Re u}\to+\infty$ parts of the instanton effectively see $v(t,x)\approx v(0,x)$, which means we can use~\eqref{dtdxGPstat}, with $p$ being the asymptotic momentum of the $\text{Re }u\to\infty$ particle. We can still use~\eqref{dtdxGPstat} shortly after the $\text{Re }u\to-\infty$ particle has entered the plateau region, so 
\be
\dot{x}(\tilde{u}_0)\approx\sqrt{\alpha^2+p^2} \;,
\ee
where $\alpha:=-v(0,-\infty)$ in the general case (i.e. not just for~\eqref{vPlateau}),
and
\be
\dot{t}(u)\approx\dot{t}(\tilde{u}_0)\approx-\frac{\sqrt{1+p^2}+\alpha\sqrt{\alpha^2+p^2}}{\alpha^2-1} \;.
\ee
We also have $v[t(\tilde{u}_0,-\infty)]\approx v(0,-\infty)=-\alpha$, so~\eqref{vPlateau} becomes
\be\label{plateauApprox2}
\dot{x}(u)\approx\dot{x}(-\infty)+\dot{t}(\tilde{u}_0) v[t(u),-\infty] \;,
\ee
where
\be\label{plateauApprox3}
\dot{x}(-\infty)\approx-\frac{\alpha\sqrt{1+p^2}+\sqrt{\alpha^2+p^2}}{\alpha^2-1} \;.
\ee
Thus, given the momentum of the particle which is immediately emitted, $p$, and the height of the plateau, $\alpha=-v(0,-\infty)$, the asymptotic momentum of the particle which initially falls into the black hole is given by~\eqref{plateauApprox3}. Eq.~\eqref{plateauApprox3} is independent of $\omega$ and of the pulse shape in both the $t$ and $x$ directions. That~\eqref{plateauApprox3} is finite means that we have indeed managed to avoid the problem with the inner horizon accelerating the particle to close to the speed of light.  

However, $\dot{x}(-\infty)<0$ so the in-falling particle still escapes with positive momentum, $p^{\prime1}=-\dot{x}(-\infty)>0$, so both particles go to $x\to+\infty$. Given that $v<0$ for real $t$ and $x$, so that in the analog case $v$ can be thought of as a fluid flowing in the $x\to-\infty$ direction, it might at first seem counterintuitive that the in-falling particle should change direction and go to $x\to+\infty$ rather $x\to-\infty$. To understand this, consider the following. Imagine that you are walking on a moving walkway (e.g. at an airport). The surface is moving at a velocity $v(t)<0$ and you are trying to walk in the opposite direction at a speed $w=\text{const.}>0$ relative to the surface of the walkway. When $|v(t)|>w$ you are moving backwards, but if the walkway operator gradually turns off the walkway, you will start to move forwards when $|v(t)|<w$, and for asymptotic times you will walk with a velocity $w$. But why were you walking against the direction of the walkway? The answer is given by~\eqref{dtdxGPstat}. In order for one of the particles to escape to $x\to+\infty$, we take $\epsilon=1$. We have seen that the complex instantons can go around branch points, e.g. as for~\eqref{AGPstatBranch}, so one might wonder whether one would not be able to go around the branch point in~\eqref{dtdxGPstat} so that $\dot{x}(\tilde{u}_0)$ would have been $-\sqrt{\alpha^2+p^2}$ rather than $+\sqrt{\alpha^2+p^2}$. However, we are thinking about the fate of the in-falling particle in the Hawking case~\eqref{Akappa}, where the instanton can be chosen to go along the real axis except for a small semi-cirlce around the pole, so the instanton would stay on the same branch of the square root and hence $\dot{x}(\tilde{u}_0)=+\sqrt{\alpha^2+p^2}$.              

In deriving~\eqref{plateauApprox3}, we have assumed that $v(t,x)\approx v(t,-\infty)$ for all $u$ once the particle has entered the plateau region ($r\lesssim-3$ in the above example). But since the $\text{Re }u\to-\infty$ particle turns around, we have to check that it does not have time to reach the horizon at $x=0$ before $v$ has become negligible. To check this, we first integrate~\eqref{plateauApprox2},
\be
x(t)\approx\frac{1}{\omega}\left[x'(\infty)\omega t-\alpha f(\omega t)\right]+c \;,
\ee
where $v(t,-\infty)=:-\alpha f(\omega t)$ (so $f(y)=\tanh y$ for~\eqref{vPlateau}), 
\be
x'(\infty)=\frac{\dot{x}(-\infty)}{\dot{t}(-\infty)}\approx\frac{\alpha\sqrt{1+p^2}+\sqrt{\alpha^2+p^2}}{\sqrt{1+p^2}+\alpha\sqrt{\alpha^2+p^2}}
\ee
and $c$ is a constant which we can neglect since $x=\mathcal{O}(\omega^0)$ at the start of the plateau. Choose $t_0$ so that $f'(\omega t)<\epsilon$ for $t>t_0$ for some $\epsilon\ll1$. For~\eqref{vPlateau} we could choose e.g. $\omega t_0\sim4$, so that $\text{sech}^2(\omega t_0)=\mathcal{O}(10^{-3})$, and $f(\omega t_0)\approx f(\infty)=1$. Demanding $x(t_0)<0$ gives
\be\label{plateauCondition}
\omega t_0 x'(\infty)-\alpha f(\infty)<0
\ee
as a validity condition for~\eqref{plateauApprox3}. For $\alpha\gg1$ we have $x'(\infty)=\mathcal{O}(\alpha^{-1})$, so~\eqref{plateauCondition} should hold. Close to the threshold $\alpha\gtrsim1$, $x'(\infty)\lesssim1$ and then, assuming natural values, \eqref{plateauCondition} would not be satisfied. For~\eqref{vPlateau} and $p=0.3$, \eqref{plateauCondition} is satisfied for $\alpha\gtrsim2.7$, and in Fig.~\ref{fig:plateau} we can see that the approximation is indeed valid for $\alpha=3$.

\section{Outlook}

We have shown how to use open worldline instantons to study pair production by gravitational fields that depend on both space and time. The fields we have considered have mainly been chosen in order to illustrate the methods. An obvious next step is therefore to consider more realistic fields. We have seen that finding instantons involves some manual work, because one has to find suitable $u$ contours in a complex plane where branch points and singularities move around as one changes the values of the field parameters or the momenta. The fact that there can be several relevant and non-equivalent instantons, which go around the branch points in different ways, motivates further research to try to find a systematic approach for finding all the relevant instantons. It could also be relevant to study instantons for which both ends of the $u$ contour go to $\text{Re }u\to+\infty$.   

In this paper we have focused on the exponential part of the pair-production probability. The pre-exponential part is obtained by expanding around the instanton/saddle points to quadratic order and performing the resulting Gaussian path integrals, which gives the functional determinant of the functional Hessian matrix. A powerfull method for computing the determinant, without having to compute the eigenvalues, is the Gelfand-Yaglom method, as shown in~\cite{Dunne:2006st} for closed instantons for Schwinger pair production. This has been generalized to open instantons and to other QED processes in~\cite{DegliEsposti:2022yqw,DegliEsposti:2023qqu,DegliEsposti:2023fbv}. It is also possible to explicitly compute the eigenvalues of the Hessian and obtain the determinant from the product of the eigenvalues~\cite{DegliEsposti:2024upq}. We leave the generalization to the gravitational case for the future.   

It was shown in~\cite{DegliEsposti:2023fbv,DegliEsposti:2021its,DegliEsposti:2024rjw} how to use open instantons to study pair production in electric fields, stimulated by a photon or an electron in the initial state.  
One could similarly consider gravitational pair production stimulated by some high energy initial particle, e.g. particle$+$black hole$\to$stimulated Hawking$+$black hole.

In the electromagnetic case, (closed) instantons have also been used for production of magnetic monopoles~\cite{Affleck:1981bma,Gould:2019myj,Gould:2017fve}, which are strongly coupled, $\alpha_m\gg1$. One potential project could be to consider magnetic monopoles produced by black holes.

\begin{figure}
    \centering
    \includegraphics[width=.39\linewidth]{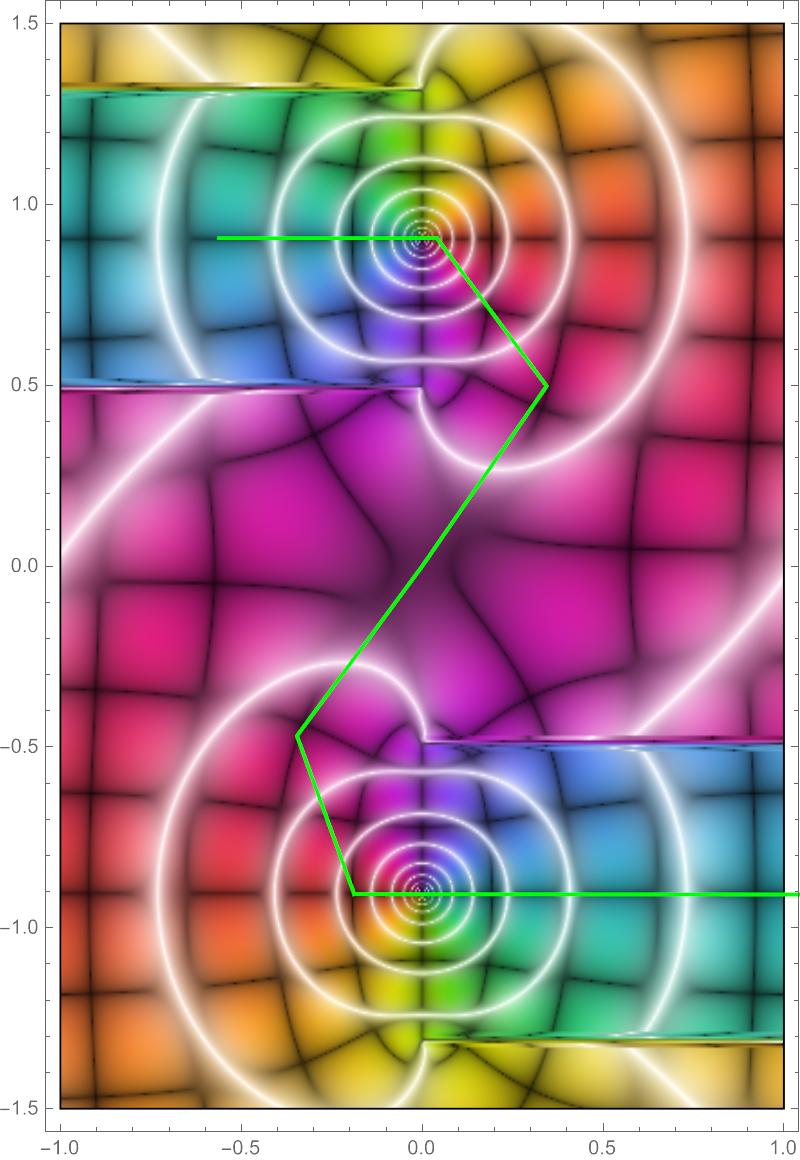}
    \includegraphics[width=.59\linewidth]{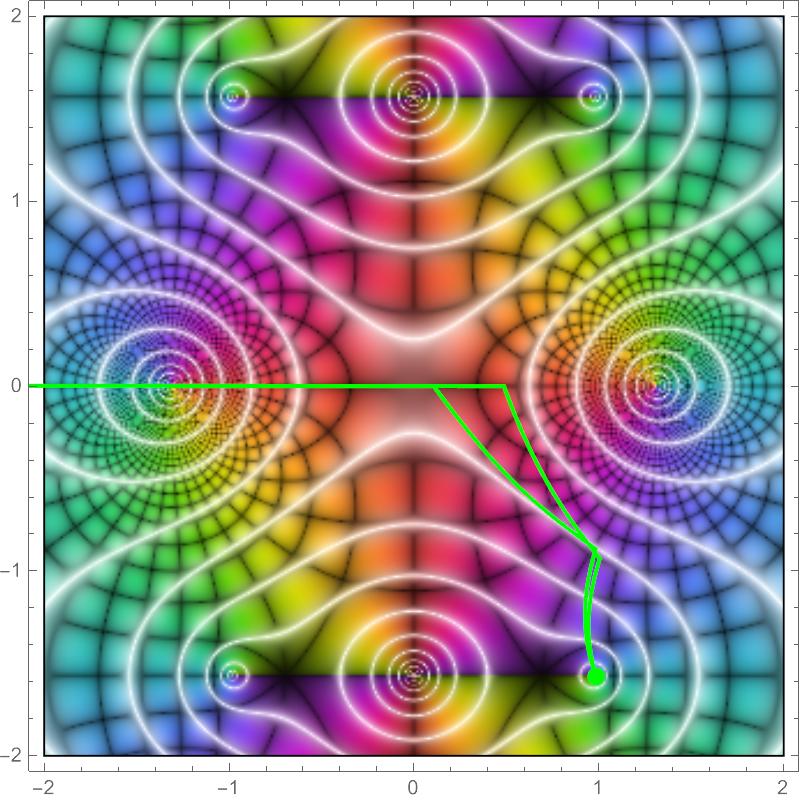}\\
    \includegraphics[width=\linewidth]{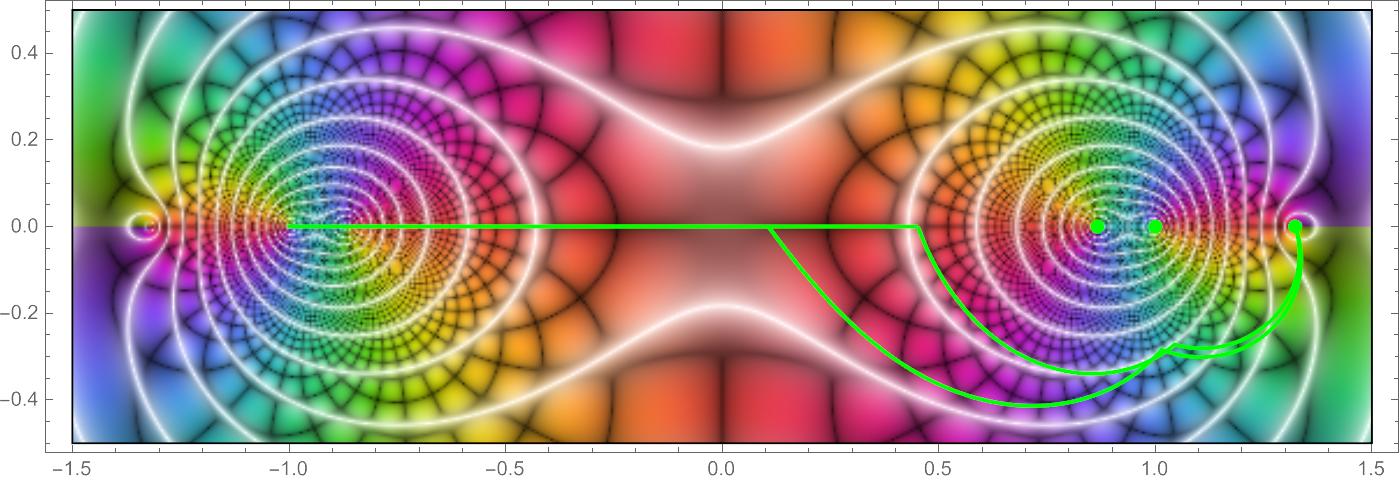}
    \caption{Instanton for~\eqref{exactAv} with $\alpha=2$ and $p^1=p_s^1\approx-1.73$. First plot: $x(u)$ in the complex $u$ plane; second plot: integrand in~\eqref{AGPstatBranch} in the complex $x$ plane; third plot: integrand in~\eqref{Aintf} in complex $f=\tanh(kx)$ plane.}
    \label{fig:exactAplot}
\end{figure}

\acknowledgements

Special thanks go to Gianluca Degli Esposti for inspiring discussions in the QED context.
We thank Karthik Rajeev for several useful discussions and for literature suggestions.
We also thank Anton Ilderton, Christian Schubert and Michael Bradley for useful discussions. We thank A.~Ilderton and K.~Rajeev for sharing a draft of~\cite{IldertonRajeev}, in which they obtain wavefunctions using open worldlines for Hawking radiation in a Vaidya metric.
G. T. acknowledges support from the Swedish Research Council,
Contract No. 2020-04327.

\appendix

\section{An analytical comparison}

Both the Hakwing result~\eqref{Akappa} and the de Sitter result (first term in~\eqref{A1and3}) are very simple, i.e. $\mathcal{A}$ is a trivial function of the parameters. In this appendix, we will compare with an example where $\mathcal{A}$ is nontrivial but still has an analytical expression. Consider metrics of the form
\be
\ud s^2=[1+\alpha F(\omega t)]^2(\ud t^2-\ud x^2) \;,
\ee
where $F$ is some dimensionless, order-one function, e.g. a Sauter pulse $F(v)=\text{sech}^2(v)$ or a Lorentzian pulse $F(v)=1/(1+v^2)$. After changing integration variable in~\eqref{generalExpFin} from proper time $u$ to coordinate time $t$, we find
\be
\begin{split}
\mathcal{A}&=2\text{Im}\int_{+\infty}^{+\infty}\ud t\frac{\alpha t(1+\alpha F)\partial_t F}{\sqrt{p^2+(1+\alpha F)^2}}\\
&=-2\text{Im}\int_{+\infty}^{+\infty}\ud t\sqrt{p^2+(1+\alpha F)^2} \;,
\end{split}
\ee
where the contour goes around a branch point, $t_B$, of the square root.

For $\alpha\ll1$, $F(t_B)$ has to be large. For fields such as the Sauter or the Lorentzian pulses, which have poles at $t=i\nu$ ($\nu_\text{Sauter}=\pi/2$ and $\nu_\text{Lorentz}=1$), $t_B$ will approach $i\nu$. We can choose an L-shaped contour, where the vertical parts go between $t=\text{Re }t_B\sim0$ and $t=t_B\sim i\nu$, and the horizontal parts go between $\text{Re }t_B$ and $+\infty$. The horizontal parts do not contribute to the imaginary part. For $\alpha\ll1$, the vertical parts can be approximated using the fact that the integrand is approximately constant over most of the contour,
\be\label{perturbativeLimit}
\alpha\ll1: \qquad
\mathcal{A}\approx\frac{4\nu}{\omega}\sqrt{1+\rho^2} \;,
\ee
which agrees with the perturbative result~\eqref{AperturbativeLimit}. 

For pulses such as $F(v)=\text{sech}^2(v)$ or $F(v)=1/(1+v^2)$, the derivative of the metric has two saddle points on the real axis (a maximum and a minimum), which in general doubles the number of branch points to consider, compared to e.g. $F(v)=\tanh t$. So, to avoid dealing with multiple instantons, and to keep the normalization at $t\to+\infty$, we consider for simplicity $F(v)=(1/2)(1-\tanh t)$, which allows us to find an analytical result,
\be
\mathcal{A}=\frac{\pi}{\omega}\left[\sqrt{1+\rho^2}+\sqrt{(1+\alpha)^2+\rho^2}-|\alpha|\right] \;.
\ee
For $\alpha\ll1$ we recover~\eqref{perturbativeLimit}. For $\alpha=-1$ and $\rho>0$ we find agreement with Eq.~(3.14) in~\cite{Hashiba:2022bzi}.

\section{Analytical $\mathcal{A}$ for a static pulse}

In this section we will study the generally non-local integral in~\eqref{AGPstatBranch} for a field which allows us to find $\mathcal{A}$ analytically, namely
\be\label{exactAv}
v(x)=-\alpha\text{sech }kx =-\alpha\sqrt{1-f^2(kx)} \;,
\ee
where $f(y)=\tanh y$. This is one example in a class of fields implicitly defined by
\be
f'(y)=[1-f^2(y)]^\beta \;,
\ee
where $\beta$ is a constant, which has been used to study similar integrals in~\cite{Gies:2015hia,Dinu:2018efz}. Changing variable in~\eqref{AGPstatBranch} from $x$ to $f$ gives
\be\label{Aintf}
\mathcal{A}=\frac{2}{k}\int\frac{\ud f}{(1-f^2)^\beta}\frac{\sqrt{p^2+\alpha^2(1-f^2)}}{1-\alpha^2(1-f^2)} \;.
\ee
We will focus on $\beta=1$.
Although we can integrate~\eqref{Aintf} without first finding an explicit instanton, it helps to plot instantons to see what $f$ contours are allowed. We can, without loss of generality, start at $\dot{x}(0)=0$, which corresponds to the branch point $f[kx(0)]=f_B$, where
\be
f_B=\frac{\sqrt{p^2+\alpha^2}}{\alpha} \;.
\ee
One possible contour is shown in the complex $u$, $x$ and $f$ planes in Fig.~\ref{fig:exactAplot}. We can see that this $f$ contour goes around the branch point $f_B$, but not around the poles at the horizon,
\be
f_H=\frac{\sqrt{\alpha^2-1}}{\alpha} \;,
\ee
or at infinity, $f=\pm1$. However, the contour can be deformed so that $f$ follows the real axis, except for two small semi-circles around the poles at $f_H$ and $f=1$. Since the square root changes sign, we obtain the same contribution for the parts that go toward and away from $f_B$. The integral in~\eqref{Aintf} can therefore be performed with the residue theorem. So for this particular example, the generally non-local integral becomes local. But there is a contribution from the pole at $f_H$ and also from $f=1$, while Hawking's result~\eqref{Akappa} for $p>0$ comes only from $f_H$. Combining both gives
\be\label{AexactBeta1}
\mathcal{A}=\frac{2\pi}{k}\left(\frac{\alpha\sqrt{1+p^2}}{\sqrt{\alpha^2-1}}+p\theta(-p)\right) \;,
\ee
where $\theta$ is the step function. In the upstream region, $x\to+\infty$ so $p>0$, the probability $P$ is maximized at $p=0$. If one only looks at~\eqref{Akappa}, it might look like $p=0$ is a saddle point, but it is not, due to the step function. There is no reason to expect there to be a saddle point at $p=0$, because of the fundamental difference between traveling upstream, $p>0$, and traveling downstream, $p<0$. In the downstream region there is instead a genuine and nontrivial saddle point,
\be\label{AexactBeta1saddle}
p_s=-\sqrt{\alpha^2-1}
\qquad
k\mathcal{A}(p_s)=\frac{2\pi}{\sqrt{\alpha^2-1}} \;.
\ee
  
In the limit where the two horizons converge and disappear, $\alpha\to1$, we find $\mathcal{A}\to\infty$ and hence $P\to0$. This is due to the contribution from the pole at the horizon, which moves closer and closer to the maximum of $v(x)$, i.e. $x_H\to0$ or $f_H\approx\sqrt{2(1-\alpha)}\to0$.   

If we take the opposite limit, $\alpha\gg1$, with $p=\mathcal{O}(\alpha^0)$, then $\mathcal{A}=\mathcal{O}(\alpha^0)$. However, \eqref{AexactBeta1saddle} shows that $P$ is maximized at $p=\mathcal{O}(\alpha)$, where
\be
k\mathcal{A}\approx\frac{\pi}{\alpha}\frac{(p/\alpha)^2+1}{|p/\alpha|} 
\qquad
k\mathcal{A}(p_s)\approx\frac{2\pi}{\alpha} \;.
\ee
Thus, $P(p<0)$ becomes less exponentially suppressed as $\alpha\to\infty$, so $P(p<0)$ becomes many orders of magnitude larger than $P(p>0)$. 

\section{Analytical $\mathcal{A}$ for a smooth static step}

Next we consider a smooth step,
\be\label{vxexp}
v(x)=-\frac{\alpha}{1+\beta e^{kx}} \;.
\ee
As always, $k\mathcal{A}$ is independent of $k$. We can also make a constant shift $kx\to kx-\ln\beta$ to remove $\beta$. We can perform the integral in~\eqref{AGPstatBranch} by first changing variable from $x=\ln y$ to $y$. This is another example where we can perform the integral with the residue theorem. We find
\be\label{AexpAna}
k\mathcal{A}(p<0)=2\pi\left(\frac{\alpha^2\sqrt{1+p^2}+\sqrt{\alpha^2+p^2}}{\alpha^2-1}+p\right) \;,
\ee
while Hawking's result~\eqref{Akappa} is given by
\be
k\mathcal{A}(p>0)=2\pi\frac{\alpha\sqrt{1+p^2}}{\alpha-1} \;.
\ee
We again find continuity at $p=0$, $\mathcal{A}(-\epsilon)=\mathcal{A}(+\epsilon)$, and that there is a nontrivial saddle point at $p<0$.

For $\alpha\gg1$ we find
\be
k\mathcal{A}(p<0)\approx\frac{2\pi}{\alpha}\left(\frac{1}{2\rho}+\rho+\sqrt{1+\rho^2}\right) \;,
\ee
where $\rho=-p/\alpha=\mathcal{O}(1)$. At the saddle point, $\rho_s=1/\sqrt{3}$ we have 
\be
k\mathcal{A}(p_s)\approx\frac{3^{3/2}\pi}{\alpha} \;.
\ee

One can also find $\mathcal{A}$ analytically for
\be
v(x)=\alpha(\tanh[kx]+h) \;,
\ee
for arbitrary constant $h$. For $h=-1$, $\alpha\to\alpha/2$ and $k\to k/2$ we recover~\eqref{AexpAna}.

\end{document}